\documentclass[12pt,apjpt4]{aastex}
\newcommand\etal{{\it et~al.}}
\newcommand\etals{{\it et~al.'s}}
\newcommand\teff{{\it T$_{\rm eff}$}}
\newcommand\logg{{\rm log~g}}

\newcommand\kms{{km~s$^{-1}$}} 
\newcommand\logb{{log(B/H)}}
\newcommand\logf{{log(Fe/H)}}
\newcommand\logn{{log(N/H)}}

\slugcomment{In preparation for the Astrophysical Journal}
\shortauthors{Mendel et al.}
\shorttitle{Boron in B-stars}
\received{2005 August 11}
\begin{document}
\title{Testing Rotational Mixing Predictions with New Boron Abundances 
       in Main Sequence B-type Stars}

\author{J.\,T.\,Mendel\altaffilmark{1} and K.\,A.\,Venn\altaffilmark{2}} 
\affil{Macalester College, 1600 Grand Avenue, Saint Paul, MN, 55105; venn@macalester.edu}
\author{C.\,R.\,Proffitt\altaffilmark{3}}
\affil{Science Programs, Computer Sciences Corporation,
3700 San Martin Drive, Baltimore, MD 21218; proffitt@stsci.edu }
\author{A.\,M.\, Brooks}
\affil{Astronomy Department, University of Washington, 
Box 351580, Seattle, WA, 98195; abrooks@astro.washington.edu}
\and
\author{D.\,L.\,Lambert}
\affil{McDonald Observatory, University of Texas, 1 University Station, 
Austin, TX, 78712-1083; dll@astro.as.utexas.edu}

\altaffiltext{1}{Present address:
   Centre for Astrophysics and Supercomputing, Swinburne University of Technology,
   Mail Number 31, P.O. Box 218, Hawthorn, VIC 3122, Australia; tmendel@astro.swin.edu.au}

\altaffiltext{2}{Present address:
  Department of Physics and Astronomy, University of Victoria, Elliott Building,
  3800 Finnerty Road, Victoria, BC, V8P 1A1, Canada; kvenn@uvic.ca}

\altaffiltext{3}{Also at:
   Space Telescope Science Institute, and Institute for
   Astrophysics and Computational Science, Catholic University of America}

\begin{abstract}
New boron abundances for seven main-sequence B-type stars are determined from HST STIS 
spectroscopy around the \ion{B}{3} 2066 \AA\ line.  Boron abundances provide a unique and 
critical test of stellar evolution models that include rotational mixing since boron is 
destroyed in the surface layers of stars through shallow mixing long before other elements 
are mixed from the stellar interior through deep mixing.  The stars in this study are all 
on or near the main-sequence and are members of young Galactic clusters.  They show no 
evidence of mixing with gas from H-burning layers from their CNO abundances.  Boron 
abundances range from 12+log(B/H)$\le$1.0 to 2.2.  The boron abundances are compared to 
the published values of the stellar nitrogen abundances (all have 12+log(N/H) $\le$7.8) and 
to their host cluster ages (4 to 16 Myr) to investigate the predictions from models of 
massive star evolution with rotational mixing effects.    We find that the variations in 
boron and nitrogen are generally within the range of the predictions from the stellar 
evolution models with rotation (where predictions for models with rotation rates from 0 to 
450 \kms\ and $\mu$-barriers are examined), especially given their age and mass ranges.
  
Three stars (out of 34 B-type stars with detailed boron abundance determinations), 
deviate from the model predictions, including HD\,36591, HD\,205021, and HD\,30836.  
The first two of these stars have much larger boron depletions than are predicted 
for their spectroscopic masses and very young ages, even adopting the highest rotation 
rates from the model predictions.   HD\,36591 also shows no significant nitrogen 
enhancement, as uniquely predicted by the rotating stellar evolution models.
HD\,205021, however, has a small nitrogen enrichment which could be explained 
by stellar rotation or mass transfer since it is in a binary system. 
The spectroscopic mass for the third star, HD\,30836, is marginally lower 
than expected given the rotating model predictions for its age and boron 
abundance.   This star also has no significant nitrogen enhancement, thus even
though it is in a binary system
it does not show the nitrogen enrichment expected if it has undergone mass transfer.
Therefore, the results from these three stars 
suggest that rotational mixing could be more efficient than currently modelled 
at the highest rotation rates.


\end{abstract}
\keywords{stars: abundances -- stars: evolution -- stars: rotation}

\section{Introduction}

Rotation is recognized as an important physical component in understanding the 
evolution of massive stars and yet is a theoretically challenging problem.  
Rotation affects the lifetimes, chemical yields, 
stellar evolution tracks, and the properties of supernova and compact remnants 
(Heger $\&$ Langer 2000; Maeder $\&$ Meynet 2000, 2005).   The new rotating stellar 
evolution models also address long standing problems such as the origin of the 
B[e] and WNL/Ofpe (slash) stars, the distribution of red to blue supergiants on 
the HR diagram, and the unseen post main-sequence gap predicted in all standard 
stellar evolution scenarios.  With rotation, it is also possible to explain the 
variations in the surface helium, carbon, nitrogen, and oxygen (He and CNO) 
abundances in OB stars on and near the main sequence 
(e.g., Gies \& Lambert 1992, Herrero \etal\ 1992, Cunha \& Lambert 1994, 
Dennisenkov 1994, Lyubimkov 1996, Lyubimkov \etal\ 2004).
Several of these observations have been used to constrain the various mixing
prescriptions used in the new models of massive star evolution with rotation. 
Additional observations are now necessary to test the model predictions
and provide new constraints for the transport of angular momentum and chemical
species in rotating massive stars.  One avenue of observational testing is to 
determine the helium and CNO abundances in massive stars in young star clusters 
in the Galaxy and Magellanic Clouds (Evans \etal\ 2005).  This tests the 
metallicity effects predicted by the models.  Another line of research includes 
the determination of the light element abundances at the surface of massive 
stars.  This tests the earliest stages of rotational mixing (timescales and 
mixing efficiencies), {\it before} hot gas from the stellar interior is 
observable at the surface (Fliegner \etal\ 1996, Venn \etal\ 2002). 

The abundances of the light elements, lithium, beryllium, and boron (LiBeB) 
are known to be sensitive to rotational mixing in both low and high mass stars.  
LiBeB is destroyed on exposure to protons at temperatures too low for H-burning 
to have occurred ($\le$6x10$^6$~K), therefore even shallow mixing at the stellar 
surface induced by rotation can lead to LiBeB depletions.  In low mass stars, 
variations in the surface abundances of all three elements have been traced to 
rotational mixing (e.g., Boesgaard $\&$ Lavery 1986; Pinsonneault 1997).  
In high 
mass stars, only boron\footnote{There are no readily accessible 
spectral lines of Li in hot stars, and the resonance lines of \ion{Be}{2} 
near 3130\AA\ are predicted to be very weak in B-type stars, and they occur 
at the atmospheric cutoff, thus these spectra would be difficult to access 
from ground based telescopes.   Currently, there are no published observations
of the \ion{Be}{2} lines in B-stars.}
has been available.   Spectroscopy of the \ion{B}{3} feature 
at $\lambda$2066 using the International Ultraviolet Explorer (IUE) 
archived spectra or the Hubble Space Telescope (HST) 
Space Telescope Imaging Spectrograph (STIS) 
or Goddard High Resolution Spectrograph (GHRS) have made boron
abundance determinations possible in B-type stars. 
Significant variations in the boron abundances in hot, massive stars 
have been observed (e.g., Proffitt \& Quigley 2001, Venn \etal\ 2002).

Boron depletions in B-type stars are predicted to be associated with nitrogen 
enhancements.  This is because mechanisms that can deplete boron at the surface 
of a star will also mix the surface with CN-cycled gas, i.e., gas from H-burning
layers where the CN-cycle has converted carbon into nitrogen.
This is true whether boron is destroyed through rotational mixing (where hot
CN-cycled gas from the interior is mixed to the surface) or through binary
mass transfer (where CN-cycled gas is deposited on the surface of the star; 
Wellstein 2001).   Simple initial abundance variations and mass loss 
from B-type stars are ruled out (see Venn \etal\ 2002).   However, there is
one {\it unique} characteristic of the rotational mixing scenario that makes
for an interesting and exciting new observational constraint.   
Boron depletion occurs {\it before} nitrogen enhancement.  This is because 
rotational mixing taps the surface boron-free layers first, then subsequent 
deeper mixing can penetrate layers where H-burning has occured via the CN-cycle,
converting carbon into nitrogen.  Thus, the initial phases of rotational mixing 
in stars is revealed by depletions of boron, followed only {\it later} by 
nitrogen-enrichment and carbon-depletion, 
a different signature from binary mass transfer.

Some B-type stars with boron depletions, but no nitrogen enrichments, were
found by Venn \etal\ (2002) and intrepreted as concrete and unambiguous evidence 
for rotational mixing on the main sequence.  However, two stars with masses of 
12-13~$M_{\sun}$ showed uncharacteristically large boron depletions.  These two 
stars required models with higher masses (20 M$_\odot$) and the highest rotation 
rates ($\sim$450~\kms) to reproduce the boron depletions (from Heger \& Langer 
2000 models).  Since very few stars are expected to rotate at these high speeds, 
and because of the mass discrepancy, this suggested that rotational mixing may be 
even more efficient than currently predicted.

Presently, 32 solar neighborhood (see Table~9 in Venn \etal\ 2002) 
and two stars in the Small Magellanic Cloud (Brooks \etal\ 2002)
B-type stars have boron abundance determinations.  Of these, only nine
have been determined from high quality HST STIS or GHRS 
spectra of the relatively unblended \ion{B}{3} 2066 line. 
In this paper, we present new boron abundances for seven additional B-type 
stars from HST STIS spectroscopy.   These stars were selected from the group 
of B-type stars examined with International Ultraviolet Explorer (IUE) that may 
have low boron abundances (Proffitt \& Quigley 2001).  
They are also all in young galactic clusters (for age estimates), 
and they show no nitrogen enrichments (12+log(N/H) $\le$7.8).  
Therefore, the determination of boron abundances in these stars 
provides a unique observational test for studying the earliest phases of 
massive star rotational mixing effects.

\section{Target Selection and Observations}

Seven B-type stars were observed using HST STIS
spectroscopy of the \ion{B}{3} feature at $\lambda$2066.  
Previous analyses of the target stars (Gies $\&$ Lambert 1992; Cunha $\&$ Lambert 1994) 
show that the targets have normal, unenriched nitrogen abundances in their atmospheres.  
Analyses of IUE spectra by Proffitt $\&$ Quigley (2001) indicate that these targets have 
very weak or absent \ion{B}{3} lines, however the resolution of IUE and signal-to-noise 
of some of these spectra make quantitative boron abundances highly uncertain.  These 7 
stars were also selected from young galactic OB associations in order that effects of 
rotationally induced mixing could be examined with age.  Data considered in the 
selection of these targets is listed in Table~\ref{basics}.

Spectra for six of the seven targets were gathered using the E230M grating 
and 0.2x0.05ND slit with a neutral density filter (ND=2) to avoid the MAMA-NUV 
brightness limits.  This grating choice is sufficient to resolve the \ion{B}{3} 
line in all of our targets since the instrumental broadening of this grating is 
9~\kms\ (given our slit choice), whereas the targets have rotational velocities 
$\le$40~\kms.  To attain the best possible signal-to-noise spectra, three 
different configurations of the E230M grating were used, each with a slightly 
different central wavelength ($\lambda_c$ of 1978, 2124, and 2269~\AA) 
so that the \ion{B}{3} line landed on different regions of the MAMA detector 
and subsequent combinations could better reduce pixel noise variations.  
Echelle observations with the E230M grating included 40 orders; only those 
orders in the desired wavelength region ($\lambda\lambda$2044-2145) were 
extracted and combined using standard IRAF\footnote{IRAF is distributed by 
the National Optical Astronomy Observatories, which is operated by the 
Association of Universities for Research in Astronomy, Inc., under cooperative 
agreement with the National Science Foundation} packages.  Spectra were then 
normalized using a low order spline3 polynomial; wavelengths listed here are
in air and at rest.  Exposure 
times, central wavelengths, and observation dates are listed in Table~\ref{obs}, 
as well as the signal-to-noise ratio of the final extracted spectra. 

The reductions for two stars, HD\,36285 and HD\,214993, were conducted using 
procedures different from those described above.  HD\,36285 (V=6.32) was faint 
enough to be observed with the higher resolution E230H grating (31x0.05NDA slit), 
and HD\,214993 is a known variable star.   The observations and reductions for 
these two stars are described below.

\subsection{HD\,36285}

Observations of HD\,36285 were done using the STIS E230H grating at the $\lambda$2013 central 
wavelength setting.  This star was too bright to be observed with any of the clear STIS 
apertures, but since we desired very high S/N in order to study the $^{10}$B/$^{11}$B isotope 
ratio (Proffitt \etal, in progress), we adopted the unsupported 31X0.05NDA slit, which has 
an attenuation factor of about 4.  The global count rates were 1.2x10$^5$ to 1.5x10$^5$~cnts/s 
per observation, which is modestly under the 2x10$^5$~cnts/s upper limit allowed for the 
NUV-MAMA detector.  Using a long slit and dithering our subexposures in the cross-dispersion 
direction along the length of the aperture allowed for higher S/N.  Nine different dither 
positions with a spacing between positions of about 0.15" (or about 5 pixels) were used. 
This averages any fixed pattern noise of flat-fielding errors, and removes any bad pixels or 
detector defects.

Since the 31X0.05NDA aperture is not a supported aperture, the standard reference files 
supplied by STScI cannot be used to produce extracted and flux calibrated 1D spectra.  While 
the measured throughput curve of the 31X0.05NDA aperture was used for calculating the 
scattered light background, we adopted the parameters supplied for the 0.3x0.05ND aperture. 
In addition, for some of the more extreme dither positions, the reference file which 
specifies the default locations of spectral orders on the detector had to be edited so 
that the calstis software could correctly identify each order.

In principle, the use of a long slit with the echelle could increase the amount of 
interorder scattered light and cause an error in the background subtraction.  There 
are some modest S/N (15 to 20) E230H observations of HD~39060 ($\beta$ Pic) which 
were used to check for such effects (Program GO-7512, PI Lagrange; see
Roberge \etal\ 2000). Observation o4g002020 was taken at the 2013 
central wavelength setting with the 0.1X0.03 aperture, while o4g002060 was taken at 
2263 with the 31X0.05NDA.   Four echelle orders  between 2125 and 2155~\AA\, are 
included in both observations. Uncertainties in the blaze function alignment at 
different central wavelength settings, and possible vignetting caused by the different 
locations of the orders on the detector in the two observations prevent an exact 
comparison, but when we reduce the 31X0.05NDA aperture observations of HD~39060 in 
the same way as for the HD\,36285 data, these observations show good agreement for 
the extracted fluxes and suggest that any error in the background subtraction amounts 
to no more than 4$\%$ of the mean flux level.  These observations also suggest that 
near this wavelength, the throughput of the 31X0.05DA slit is about 20$\%$ lower than 
its tabulated value.

\subsection{HD\,214993}

HD\,214993 (12 Lac) is a complex multi-mode $\beta$ Cephei variable, with at least 
six distinct pulsational frequencies (Jerzykiewicz 1978) with periods between 0.095 and 
0.236 days.  The interaction of the different modes leads to considerable variation in both 
the amplitude of the pulsation and in observed profiles of absorbtion lines.  Periods and 
amplitudes of these modes were measured by Mathias \etal\ (1994) using the \ion{Si}{3} triplet 
at 4553, 4568, and 4575 \AA. He fit the velocity curves with a sum of sinusoidal functions, 
although this fit does not perfectly track the velocity curve, possibly due to significant 
non-linearities in the pulsation amplitudes.

\begin{figure}[ht] 
\epsscale{1}
\plotone{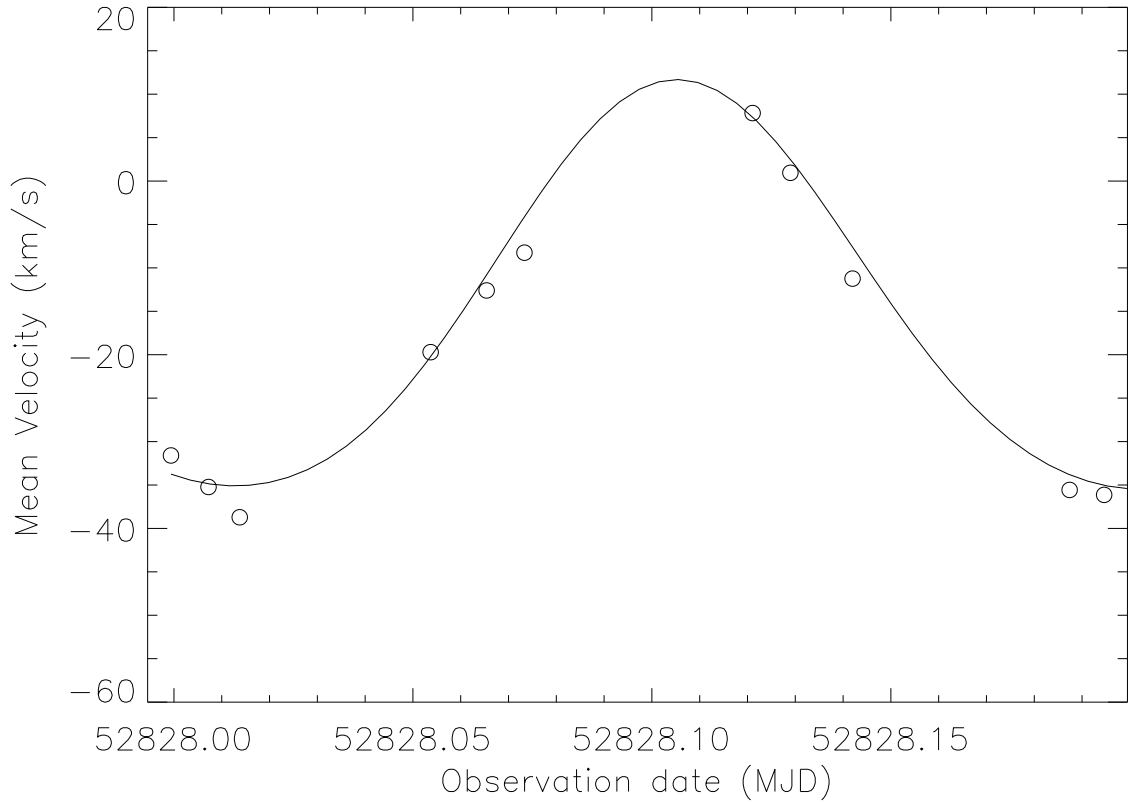}
\figcaption{Measured radial velocity of each STIS subexposure of the $\beta$-Cephei star 12~Lac (HD\,214993) is plotted as a function of the obervation start time. We have overplotted a velocity curve that adopts the pulsational frequencies and amplitudes found Mathias \etal\ (1994), but have adjusted the phases of each of the six pulsation modes to optimize the fit. \label{vel_12lac}} 
\end{figure}

Our 11 subexposures of this star spanned about 0.195 d, or about 1 full cycle of 
the largest amplitude mode (0.193 d). We measured velocity shifts for each of the individual 
sub-exposures by cross-correlating the data in order 99 ($\approx$ 2045--2080 \AA) with our 
adopted synthetic spectrum for this star.  The strong \ion{Zn}{2} ISM line was excluded from this 
cross correlation. The measured velocities are plotted vs.\ observation time in 
Fig.~\ref{vel_12lac}.

The 14 year span between the Mathias \etal\ observations and our STIS data is much 
too long to allow us to predict the phases of individual modes. However, for illustrative 
purposes we have overplotted a velocity curve that adopts Mathias \etals\ mode frequencies, 
amplitudes, and system velocity, but with phases for the individual modes arbitrarily 
adjusted to fit our observed data. The rms residuals of this fit are only about 2~\kms.

\begin{figure}[ht] 
\epsscale{1}
\plotone{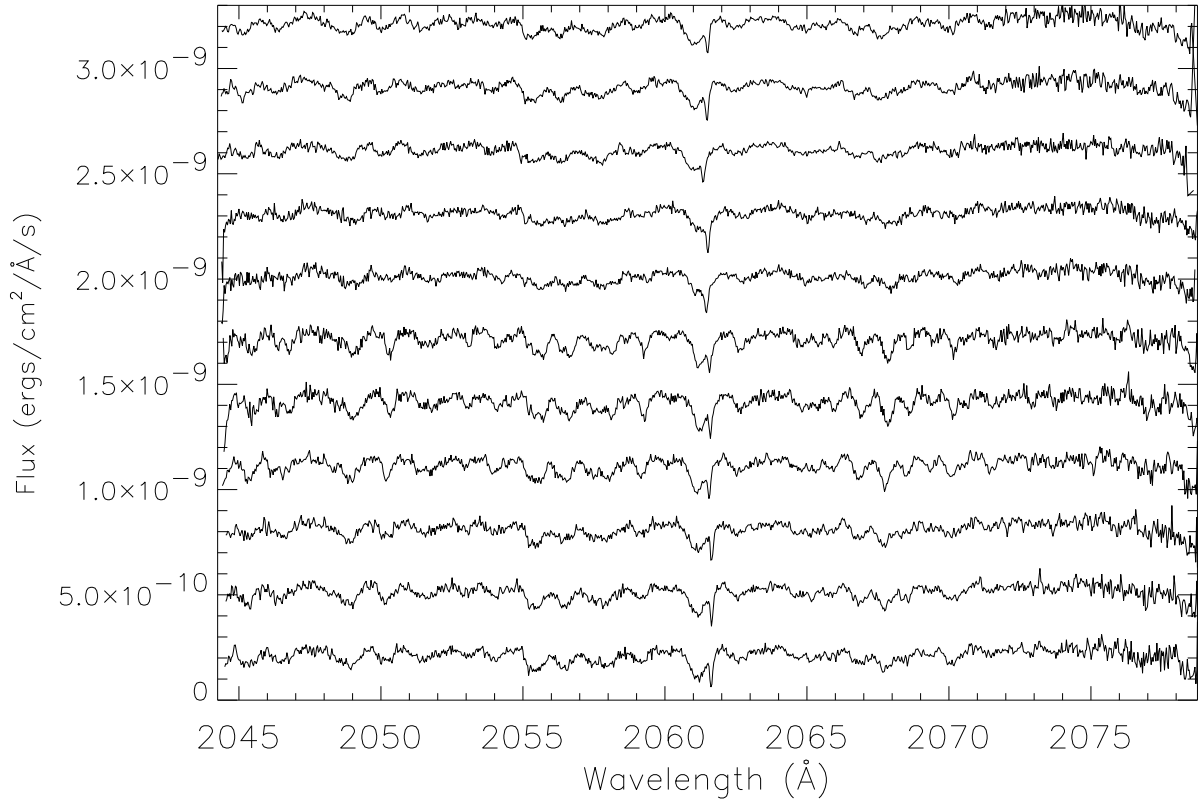}
\figcaption{We show the individual spectra from order 99 of the STIS E230M grating for each individual subexposure of 12 Lac (HD\,214993). The first subexposure is plotted at the bottom of the figure, and each of the subsequent spectra is offset by $3\times10^{-10}$ flux units. The spectra are aligned to the laboratory wavelength scale in air.\label{spec_12lac}}  
\end{figure}

\begin{figure}[ht] 
\epsscale{1}
\plotone{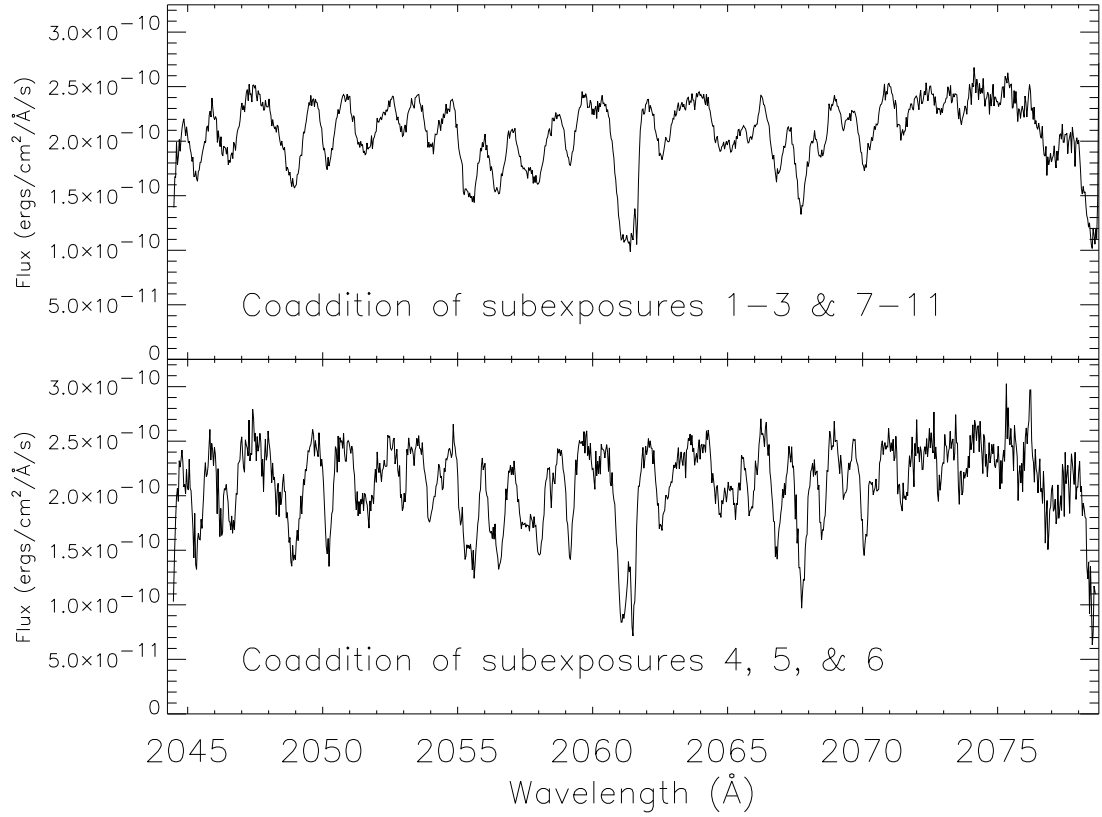}
\figcaption{Coadded spectra for the three most narrow-lined observations of 12~Lac (bottom panel) are compared to the coaddition of the data for other phases.  Each subexposure was shifted to the laboratory wavelength scale before coadding. Wavelengths are given in air.\label{coadd_12lac}}   
\end{figure}

In Fig.~\ref{spec_12lac} we show for each subexposure the observed spectrum in 
echelle order 99.  For our analysis we only used the data from subexposures 4, 5, and 6 
which show the narrowest line profiles, which were taken when the star was near its maximum 
radius. We compare the coadded data for these subexposures to the coaddition of the other 
subexposures in Fig.~\ref{coadd_12lac}. Note that the optical analysis of Gies $\&$ 
Lambert (1992) did not comment on the pulsational phase at which their observations 
were taken. This may introduce some additional uncertainties into their results for this star.

\section{Abundance Analyses}

Elemental abundances have been determined from LTE spectrum syntheses and ATLAS9 model 
atmospheres (Kurucz 1979, 1989).  Solar metallicity models were used throughout.  The 
stellar temperature and gravity values were adopted from the literature; see Table 
\ref{param2}.  For HD\,36285, the temperature is from Cunha $\&$ Lambert (1994; CL94).  
For all other targets, temperatures are from Gies \& Lambert (1992; GL92) scaled down by 
3.4$\%$ to bring them into agreement with the CL94 temperature scale.    

LTE spectrum syntheses were constructed using the program LINFOR\footnote{LINFOR was 
originally developed by H. Holweger, W. Steffen, and W. Steenbock at Kiel University.  
It has been upgraded and maintained by M. Lemke, with additional modifications by N. 
Przybilla}.  The spectral line list covers the region from 2044 to 2145 \AA, and was 
adopted from Brooks \etal\ (2002).  This line list originated from the Kurucz (1993; 
CD-18) line list, including light elements, all lines in the iron-group, and heavy 
elements up to barium, and through the third ionization states.   The line list was 
updated to include new wavelengths and/or oscillator strengths for $\sim$180 \ion{Fe}{3} 
lines (Proffitt \etal\ 1999; Ekberg 1993).   Brooks \etal\ also included some fine 
adjustments to iron-group features after an examination of the synthesis of their sharpest 
lined stars; this included 5 \ion{Mn}{3} lines and 26 \ion{Fe}{3} lines.   The final 
linelist includes 6685 features, although many are negligible contributors.   Atomic 
data for the \ion{B}{3} 2$s^{\rm 2}$S - 2$p^{\rm 2}$P resonance doublet with lines at 
2065.8~\AA\ and 2067.3~\AA\ are taken from Proffitt \etal\ (1999).  

The weaker \ion{B}{3} line at 2067.3 \AA\ is blended with a strong \ion{Fe}{3} line and 
weaker \ion{Mn}{3} line, and is not suitable for boron abundance determinations.  For all 
syntheses, an isotopic ratio of $^{\rm 11}$B/$^{\rm 10}$B=4.0 is adopted, the solar system 
ratio (Zhai \& Sahw 1994, Shima 1963) which is consistent with the values determined by 
Proffitt \etal\ (1999) from their line profile analysis of two sharp-lined B-type stars.

As discussed by Proffitt $\&$ Quigley (2001), interstellar (IS) lines must be addressed, 
particularly due to the capacity of one \ion{Cr}{2} IS line at $\lambda$2065.50 to be 
shifted into a blend with the boron feature.  All 
of our stars have been examined for IS lines.  In no instance does an IS line interfere 
with the \ion{B}{3} features.  All four of the IS lines near the \ion{B}{3} feature are 
only observable in one star, HD\,22951 (see Fig.~\ref{highres}).

\subsection{Spectrum Syntheses: I. Line by Line Analysis \label{linebyline}}

For five of the seven targeted stars, the S/N of the reduced spectra and their broadening 
parameters (sharp lines) made it possible to identify and fit individual stellar features 
for a line-by-line analysis.  The wavelengths and abundances from these (mostly iron-group)
features are listed in Table~\ref{mabund}.  Abundance results are listed relative to meteoritic 
values from Asplund, Grevesse \& Sauval (2005), 
with 12+log(Fe/H, Mn/H, Ni/H, Cr,H, Co/H) =(7.45, 5.47, 6.19, 5.64, 4.82).  
Not all iron-group features were used in the final iron-group abundances 
listed in Table~\ref{results};
due to poor atomic data in the UV and/or unrecognized blends, we iterated such that the 
average value includes only features within 2$\sigma$ of the mean.   

\begin{figure}[ht]
\epsscale{1}
\plotone{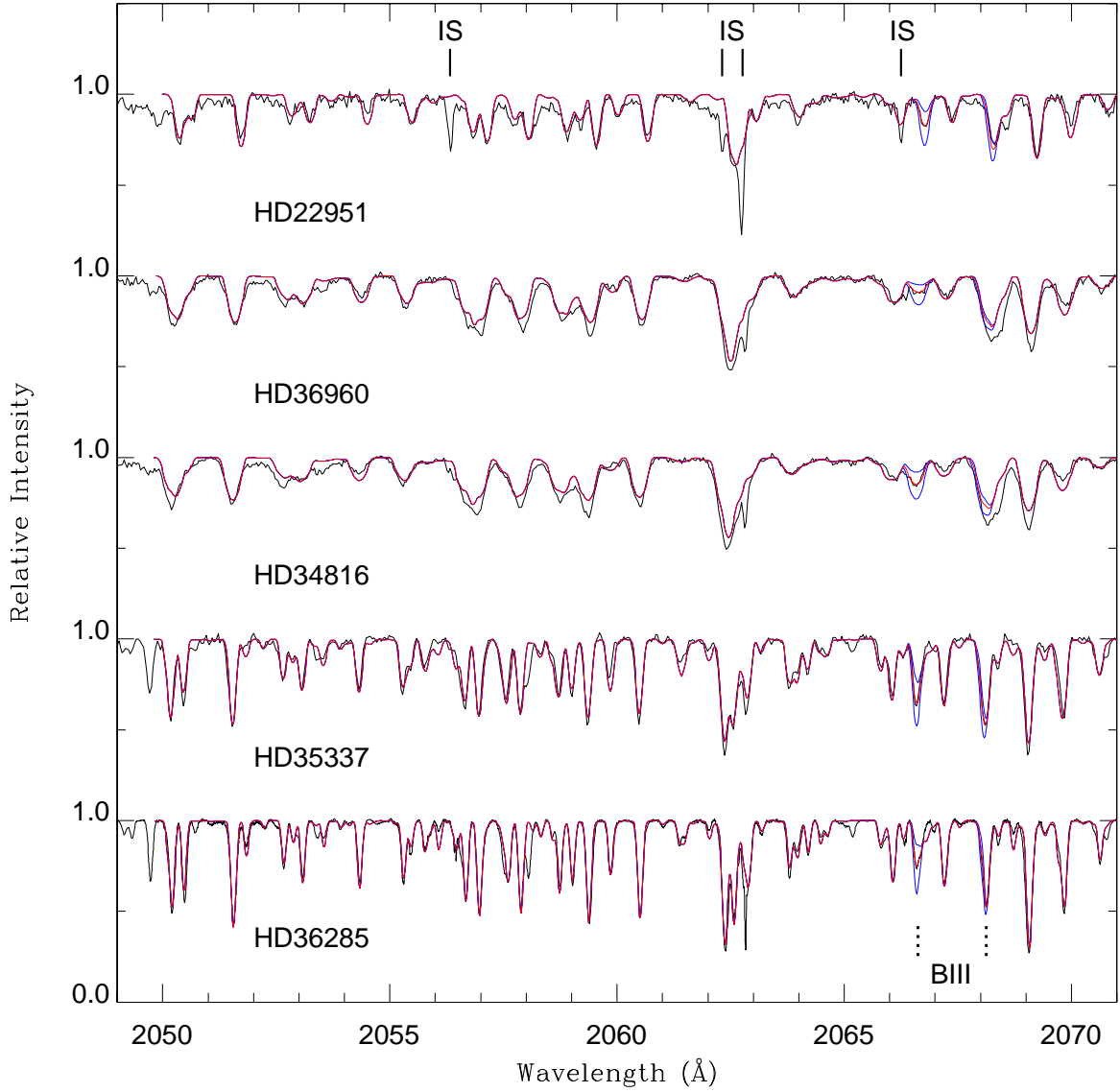}
\figcaption{Best-fit metallicites and boron abundances (red) are compared to the
spectra of five stars (black) with sufficiently high signal-to-noise spectra for 
a line-by-line analysis.  Best-fit boron abundances $\pm$0.40~ dex are shown (blue)
for comparison.  Four interstellar lines are marked above the figures, while the 
boron lines are indicated below.\label{highres}} 
\end{figure}

Microturbulence values were calculated in this line-by-line analysis by requiring that 
iron-group abundances match for both strong and weak features 
(i.e., the features in Table~\ref{mabund} within 2$\sigma$ of the mean).  
This analysis yields microturblence 
results significantly lower than ones determined using optical observations of these same stars 
(Gies $\&$ Lambert; 1992).  This same trend has been noted by both Venn \etal\ (1996)  and 
Cunha \etal\ (1997),  and is likely a reflection of differences in the atmospheric structure
between the formation of the UV and optical spectral regions.  
Macroturbulence was taken as the 
instrumental broadening (9~\kms\ for the E230M grating and 4~\kms\ for the E230H grating).  Values 
for rotational broadening and radial velocities were calculated via line profile fitting for each 
star from the iron group features in Table~\ref{mabund}.  Spectrum syntheses in 
Fig.~\ref{highres} show the best fit iron abundance.   

Boron abundances were determined from best fit syntheses of the $\lambda$2065.8 \ion{B}{3} 
line, even though it is blended with a weak \ion{Mn}{3} line at $\lambda$2065.9.  The 
measurement of this \ion{Mn}{3} line has been shown to be in good agreement with the overall 
iron-group abundance (Venn \etal\ 2002 = V02), thus the mean Fe group abundance was adopted for the Mn abundance 
prior to the boron abundance calculations.  The best fit boron abundances, and syntheses with 
[B/H]=$\pm$0.4 dex, are shown in Fig.~\ref{highres}.  The broadening parameters, and LTE 
iron-group and boron abundances are summarized in Table~\ref{results}. 

In this analyses, all of the target stars result in lower iron-group abundances 
than anticipated for objects in the Orion star forming region or the 
solar neighborhood.  Cunha, Smith \& Lambert 
(1998) examined F and G-type stars in the Orion complex and found iron abundances 
from optical observations of \ion{Fe}{1} and \ion{Fe}{2} lines that were consistent 
with other stars in the solar neighborhood (\logf$\sim -0.1$). 
The fact that our analysis of B-stars results in iron-group abundances that 
are sub-solar by [Fe/H] = 0.1 to 0.4 appears to be partially due to the 
temperatures adopted.   In V02, boron and iron-group abundances were
determined using temperatures from the literature.   The boron abundances
were then scaled to a common temperature scale, but not the iron-group abundances.
If the iron-group abundances had also been scaled (e.g., reducing the Gies \&
Lambert 1992 temperatures by 3.4\%, as done in this paper from the beginning),
then those abundances would have been reduced by $\sim$0.1~dex, in better 
agreement with the results in this paper.   
The iron-group abundances are also very sensitive to microturbulence for
which there are very few constraints in our UV spectra.  In V02, we showed 
that $\Delta\xi$~=~$\mp$1~\kms\ results in $\Delta$[Fe/H]~$\sim~\pm$0.20.
If all of the microturbulence values in this paper are reduced by 1~\kms
(with no changes in temperature) then the mean iron-group abundance is 
solar.   These changes in temperature and microturbulence have much 
smaller effects on boron, $\Delta$[B/H] $\le$0.05 and 0.02, respectively
(see Section~4).
Thus, we do not expect that the low iron-group abundances found here
affect the main science in this paper, 
i.e., accurate boron abundances in B-type stars.

\subsection{Spectrum Synthesis: II. Two Exceptions}

Of the seven target stars, two had spectra that prevented a detailed line-by-line 
study.  Because only three of the subexposures for HD\,214993 could be used, this 
resulted in a lower signal-to-noise spectrum, while HD\,30836 has very broad features
due to its high $v$sin$i$ value. 

\begin{figure}[ht]
\epsscale{1}
\plotone{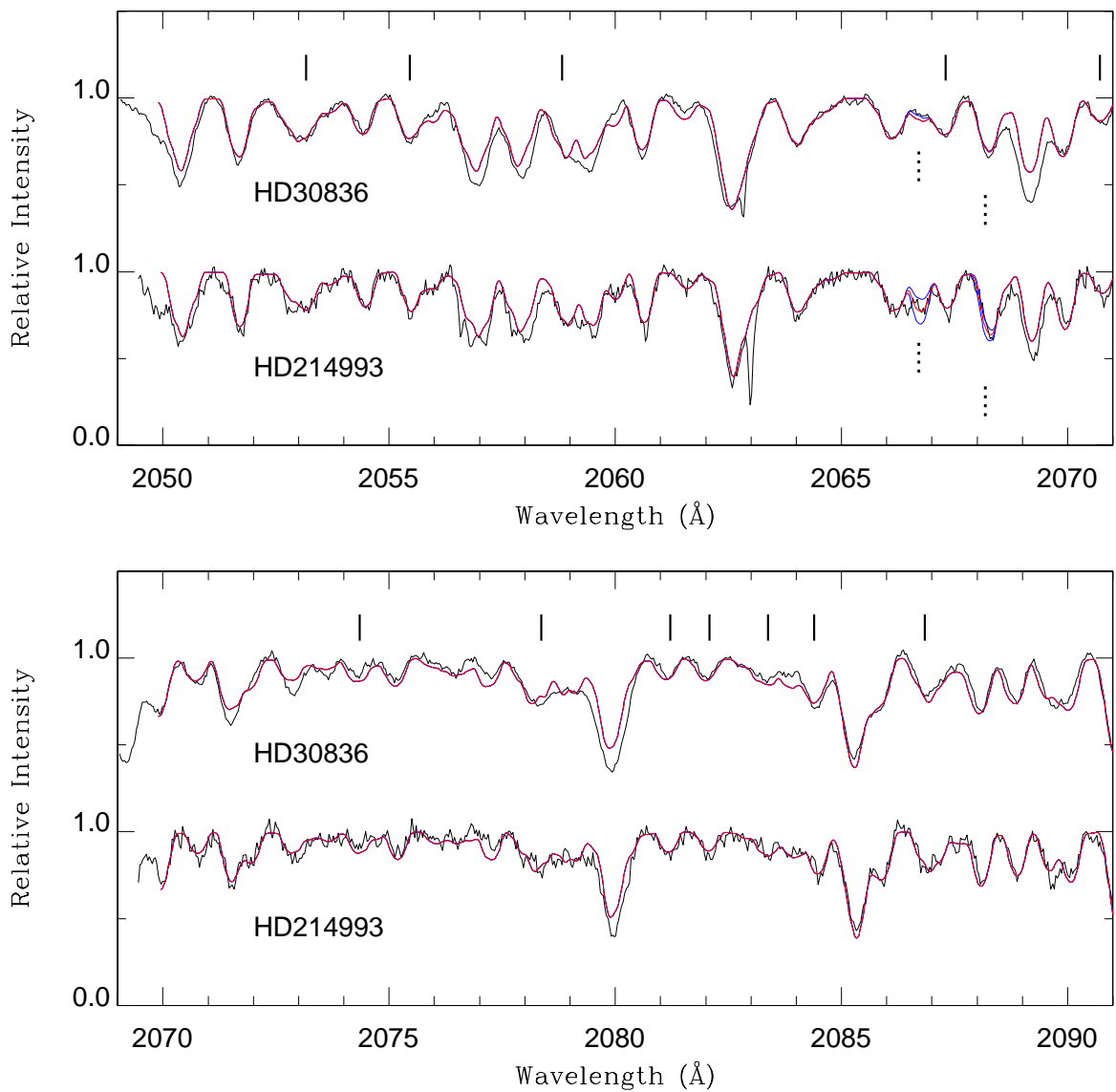}
\figcaption{Best-fit metallicities and boron abundances (red) are compared to the
spectra of two stars (black) with either lower signal-to-noise spectra (HD\,214993)
or lower gravity (HD\,30836) than the other stars in this program.  Best-fit boron
abundances $\pm$0.40~dex are shown (blue) for comparison, which also emphasizes the
upper-limit determination for HD\,30836.    
Rotational and radial velocities were determined from the most reliable iron-group
lines (in Table~\ref{mabund} and marked above the spectra).
Boron is indicated below the plots by the dotted lines.  
\label{lowres}} 
\end{figure}

Lines that were determined to be ``clean'' in the high quality spectra 
(lines that do not deviate from the mean abundance by more that 2$\sigma$ for 
the 5 stars discussed above in Section 3.1, see Table~\ref{mabund}) 
were used to find the rotational 
and radial velocities.  Using these parameters, a best metallicity and 
microturbulence was determined simultaneously (where the metallicity range
examined ranged from solar and 1/3 solar).  The best-fit spectrum syntheses 
are shown in Fig.~\ref{lowres}.  
In both cases, sub-solar metallicities were found
to fit best (see Table~\ref{results}).
For HD\,30836 and HD\,214993 the best fit iron-group abundance was adopted for 
the \ion{Mn}{3} abundance (blended with the boron line), and used to calculate 
the boron abundance or its upper limit (see Fig.~\ref{lowres}).

\subsection{Spectrum Synthesis: III. HD35299} 

\begin{figure}[ht]
\epsscale{1}
\plotone{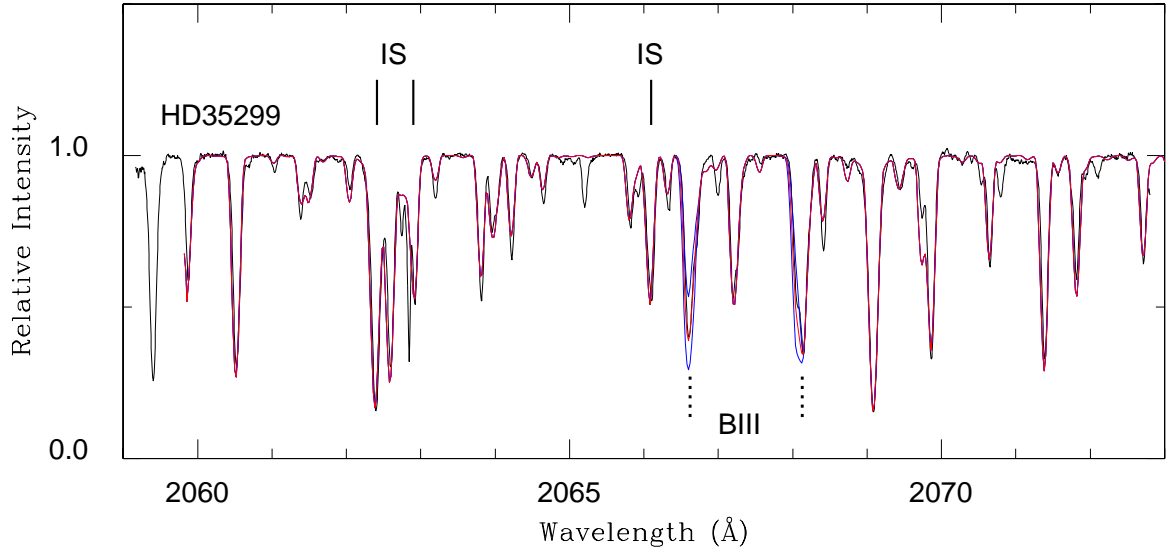}
\figcaption{The HST-GHRS spectrum of HD\,35299 (black; Proffitt \etal\ 1999) with 
best fit iron-group (\logf=7.20) and boron (\logb=2.55) spectrum synthesis (red). 
Best-fit boron  $\pm$0.40~dex is also shown (blue).  We use this star to check our
methods (see Section 3.3). 
\label{35299}} 
\end{figure}

We have synthesised the Goddard High Resolution Spectrograph (GHRS) spectrum 
of HD\,35299 near 2066\AA\ from Proffitt \etal\ (1999; see also 
Lemke \etal\ 2000) to verify that our methods and models are consistent 
with previous analyses that found a solar boron abundance for this star.  
Temperature and gravity values are from CL94.  
Macroturbulence was set to the instrumental broadening, 3.5~\kms, as 
calculated using FWHM (0.0246~\AA) from Proffitt \etal (1999).  
Rotational broadening 
(8~\kms), and radial velocities were calculated here via line profile fitting 
of the iron features deemed best in Table \ref{mabund} (although the 
spectral range is smaller $\lambda\lambda$2059-2073).  
Microturbulence was taken as the best 
fit value for each of the input abundances.
The HD\,35299 spectrum is plotted in Fig.~\ref{35299} 
and shows the best-fit iron-group (\logf=7.20) and boron 
(\logb$_{NLTE}$=2.55) abundance synthetic spectrum.
For comparison, we also show the best-fit boron abundance 
$\pm$0.4~dex.   This iron-group abundance is consistent with
the average of \logf from the two ``clean'' (as described in Section 3.1) 
iron lines that fall into the wavelength range of our observations.
This value was used to set the abundance of the 
\ion{Mn}{3} blend at $\lambda$2065.9.
This NLTE boron abundance is in good agreement with the results
from both Proffitt \etal\ 
(\logb=2.41)\footnote{This value has been adjusted to the CL94 
temperature scale and NLTE corrected using Table~8 from V02} 
and Lemke \etal\ (\logb=2.70).

\subsection{Nitrogen and Oxygen Abundances}

In Table~\ref{param2}, we have reduced the GL92 temperatures 
by 3.4\% and applied their nitrogen and oxygen abundance corrections 
for this temperature change (the $\Delta$ values in their Table~9).  
We have also applied a correction to account for differences in the LTE 
abundances determined from their use of Gold as opposed to the more 
heavily line-blanketed Kurucz model atmospheres.  
These corrections have been calculated by CL94 (their Table~10).  
Note that CL94's corrections are to the LTE abundances, and we 
assume that the same will apply to the NLTE abundances.  
For stars that have been studied only by CL94, similar 
corrections have been applied to account for the differences 
between LTE Gold and Kurucz model abundances.

\section{Boron and Iron-group Abundance Uncertainties \label{uncs}}

In this analysis, we have listed the line-to-line scatter in the iron-group abundances
in Table~\ref{mabund}.   For all other iron-group uncertainties 
(atmospheric parameters, metallicities, spectral resolution, and S/N), 
we refer to Table~7 in V02 since their targets stars, UV spectra, and 
methodology are similar to those in this paper.  Most of these atmospheric 
effects result in iron-group uncertainties $\le$0.1~dex, with the exception
of $\xi$ which is larger ($\sim$0.20 dex).
The macroturbulence and radial velocity measurements are estimated with 
uncertainties of $\pm$2~\kms\ based on the line profile fitting.  
Microturbulence is estimated as accurate to $\pm$~1~\kms\ based on line strengths.  
Overall, the iron-group uncertainties are estimated as $\Delta$\logf=0.25~dex.

To determine the uncertainties in the boron abundances, we have used similar methods
to those described in V02 and Brooks \etal\ 2002, calculating the offset in the boron abundance
due to the uncertainties in temperature and gravity (as published), macroturbulence, 
microturbulence, and radial velocities (from our iron-group analysis), and location of 
the continuum ($\pm$1\%).  The abundance of the \ion{Mn}{3} line blended 
with the \ion{B}{3} line was maintained constant during the boron uncertainty calculations.  
Brooks \etal\ showed that the error determined when Mn remains fixed is in good agreement with 
that determined by refitting the Mn line with each successive error calculation.    
The resulting boron uncertainties are listed in Table~\ref{errors}; the total error
for boron is taken as the sum of the squares of these values. 
As was found in Brooks \etal, the accuracy of the atmospheric parameters 
(temperature and gravity) is the greatest source of error in the boron abundance.

The isotopic ratio of $^{\rm 11}$B/$^{\rm 10}$B can also have a small effect on the
derived boron abundance.   In V02, a ratio of 2.0 was found to change log(B/H) by 
$\le$0.12~dex.

\subsection{The Lyubimkov \etal\ Temperature Scale}

In addition to the uncertainties in the atmospheric parameters, we 
also calculated the errors if we adopt a different temperature scale.  
Lyubimkov \etal\ (2002) report cooler temperatures for B-stars.
The iron-group and boron abundances were calculated for five of our 
stars (those that underwent the line-by-line analyses) with temperatures 
and gravities from both Lyubimkov \etal\ and GL92-3.4$\%$, and are compared 
in Table~\ref{L02param}.  Both the iron and boron abundances are lower 
by 0.02 to 0.26~dex using the Lyubimkov temperatures (and gravities).
There is no obvious correlation between these corrections and temperature, 
unlike the results for oxygen from Daflon \etal\ (2004).   The typical
correction to the iron or boron abundances is $\sim-0.1$~dex, thus
the change in temperature scale seems to have only a small effect
on our abundance determinations. 

We also note that for this test, four (out of the five) stars had values 
of \teff~ and \logg~ calculated directly by Lyubimkov \etal (2002).  
In order to correct the fifth star, HD~22951, equations 6 and 8 from 
Lyubimkov \etal\ were used to calculate the new \teff.  This choice of 
equations was based on a previous \logg=4.4 determination for HD~22951.  
This new temperature  was then used to estimate a new value for \logg\ 
(from their Fig.~13).  Values for [c$_1$], Q, and $\beta$ were obtained 
from Hauck $\&$ Mermilliod (1998).  In order to confirm our methods, 
adjusted values of \teff~ and \logg~ were calculated for HD~35337 with 
the above described method and were found to be in excellent agreement 
with the values listed in Table~1 of Lyubimkov \etal (2002). We note this 
because the iron and boron abundance corrections for HD~22951 are the
largest in Table~\ref{L02param}.

\section{Discussion \label{discussion} }

Current stellar evolution models that include rotation follow the evolution of the
angular momentum distribution and associated mixing processes in massive stars from
the pre-main sequence through core collapse.   Rotational mixing can affect stellar
surface abundances, stellar life times, and the evolution of a star across the HR
diagram, thus stellar mass estimates.   Generally, a rotating star has a lower
effective gravity, thus it acts like it has less mass at core-H ignition.   During
core-H burning, rotationally induced mixing of protons from the envelope into the
convective core and of helium from the core into the envelope will lead to higher
luminosities compared to non-rotating models.   Evolution on the main-sequence then 
depends on mass and rotation rate, as well as the efficiency of mixing in the 
upper convective core and stellar interior.

Stellar evolution models by Heger $\&$ Langer (2000) follow the changes in the 
surface abundances of boron and nitrogen as a function of time (age) for a variety 
of stellar masses (8 to 25 M$_\odot$) and initial rotation rates (0 to $\sim$450 \kms).  
These relationships are important because they show that boron is a much more sensitive 
indicator of mixing processes in B-type stars than the more commonly examined CNO abundances.  
Two sets of models are computed by Heger $\&$ Langer, with different assumptions on
the efficiency of rotational mixing in layers containing a gradient in the mean
molecular weight $\mu$ (called a $\mu$-barrier).  The differences between the two sets
of models reflects the remaining uncertainties in the theoretical description of
the rotational mixing processes.  It is interesting to note that even though the 
rotationally induced $\mu$-barrier {\it inhibits} mixing just above the core, the 
current models show that the envelope above the $\mu$-barrier is very well mixed, 
thus CN-processed gas is {\it more} evident up through the photosphere than when the
$\mu$-barrier is ignored.   The models also show the degeneracies in the stellar 
abundances (CNO and boron) with mass, age, and rotation rate.   Thus, determination of 
these stellar abundances {\it alone} cannot uniquely constrain rotational 
mixing scenarios; additional information (such as mass, rotational velocity, 
or age) of the target stars is necessary.

In this paper, we attempt to constrain both mass and age for our selection of stars
while interpreting their CNO and boron abundances.
Mass is constrained by the model atmosphere parameters; in Fig.~\ref{gvsteff2}, 
it is clear that most of our target stars are $\sim$12 M$_\odot$, ranging 
in mass from 8 to 14 M$_\odot$. 
Age is constrained by our initial
selection of stars in young clusters\footnote{The cluster ages depend on isochrone 
fitting, which in turn depend on stellar evolution models.   Standard stellar evolution
models without rotation are usually adopted when cluster ages are determined.   Rotational
mixing could have an effect on the ages of clusters, but it is not clear what the effect
on age would be.   Isochrone fitting takes into account all main-sequence (turn off) stars,
and these stars can be expected to display a variety of rotation rates.}, 
see Table~\ref{basics}. 
Unfortunately, the rotation rates themselves cannot be constrained because we are required 
to analyse sharp-lined stars for reliable boron determinations, which could be either
slowly-rotating or nearly pole-on rapidly rotating stars.   
Abt, Levato, \& Grosso's (2002) statistical survey of rotation rates of B-type stars 
in the Bright Star Catalogue found B0-2V stars with a mean $v$sin$i$ value of 
127 $\pm$8 \kms, and that the highest $v$sin$i$ value is 410 \kms, 
though these are very rare stars
(0.3\% of the sample had $v$sin$i$~$\ge$~350\kms).   
This velocity range is covered by the Heger $\&$ Langer models.
Nevertheless, with nearly all other parameters constrained, 
then we can compare the predictions from the evolution models for 
boron versus nitrogen {\it and} mass {\it and} age, 
for a range of rotation rates for our stellar sample.

%

\subsection{Initial Abundances of Boron and Nitrogen}

The meteoritic abundance of boron is adopted as the initial abundance of boron in
the B-type stars in this paper, $12+$\logb$ = 2.78\pm 0.04$ (Zhai $\&$ Shaw, 1994).  
This value is in good agreement with the new 3D solar photospheric boron abundance 
determined by Asplund, Grevesse, \& Sauval (2005), as 12+log(B/H) = $2.70\pm 0.2$, 
with corrections for departures from LTE. 
Their solar nitrogen abundance is determined from 3D analyses of NI and NH features 
(the NI result includes a small correction for NLTE), resulting in 
12+log(N/H) = 7.78 $\pm$0.06, which is in good agreement with 
other stars in the solar neighborhood.
Additionally, the interstellar boron abundance from \ion{B}{2} 1362 \AA\ 
resonance lines in diffuse clouds in the solar neighborhood has a lower limit 
of 12+log(B/H)$\ge2.4$ (Howk, Sembach, \& Savage 2000, where 
a lower limit is given since boron may be depleted onto interstellar grains).
In this paper, we assume the solar abundances are representative of the stars
in the solar neighborhood, even the younger B-type stars 
(even though one may ask why there has been so little chemical 
enrichment over the $\sim$5 Gyr lifetime of the Sun).

\subsection{Boron versus Nitrogen: Predictions and Observations}

In hot stars, the boron-nitrogen relationship is of particular interest 
because the majority of boron in the star is destroyed very early in its 
main-sequence lifetime ($\leq 10^4$~yr), leaving only the outermost layers 
to retain their initial boron abundance.  Subsequent mixing of CN-cycled 
gas towards the surface of the star through convective overshoot 
and/or semiconvection results in a downward mixing of the boron rich 
outer layers toward the stellar interior.  Because of the shallow mixing 
required for boron to be destroyed (only about 1~$M_{\sun}$ down), 
its depletion manifests itself prior to the CN-cycle nitrogen gas 
being mixed to the surface.    Predictions for the change in boron 
versus nitrogen from the Heger $\&$ Langer (2000) models are shown 
in Fig.~\ref{modbn}, where it can be seen that a clear signature of 
rotating models is that the reduction in boron is far greater than 
the increase in nitrogen.   The dotted lines in Fig.~\ref{modbn} are
from the same model (12 M$_\odot$, ZAMS V$_{\rm rot}$ = 200\kms, and
including $\mu$-barriers), but various initial boron and nitrogen 
abundances.   
Interestingly, the boron-to-nitrogen relationship shown in 
Fig.~\ref{modbn} is nearly independent of mass, rotation 
velocity, and age in the stellar models 
(see Fig.~12 in V02).

Fig.~\ref{modbn} also shows that that most of the observational data 
follows or exceeds the predicted trend.
All filled data points in Fig.~\ref{modbn} are boron determinations 
from \ion{B}{3} 2066.  Only one star has a solar boron abundance as
determined from the \ion{B}{3} lines, HD\,35299, thus we reanalysed it
here (see Section 3.3) to verify our methods.   All the other stars have
boron results from the \ion{B}{3} line below 12+log(B/H)=2.3.  For the stars 
in this paper, lower boron abundances are to be expected since targets were 
selected to have weak \ion{B}{3} lines from IUE spectra.   

\begin{figure}[ht]
\epsscale{1}
\plotone{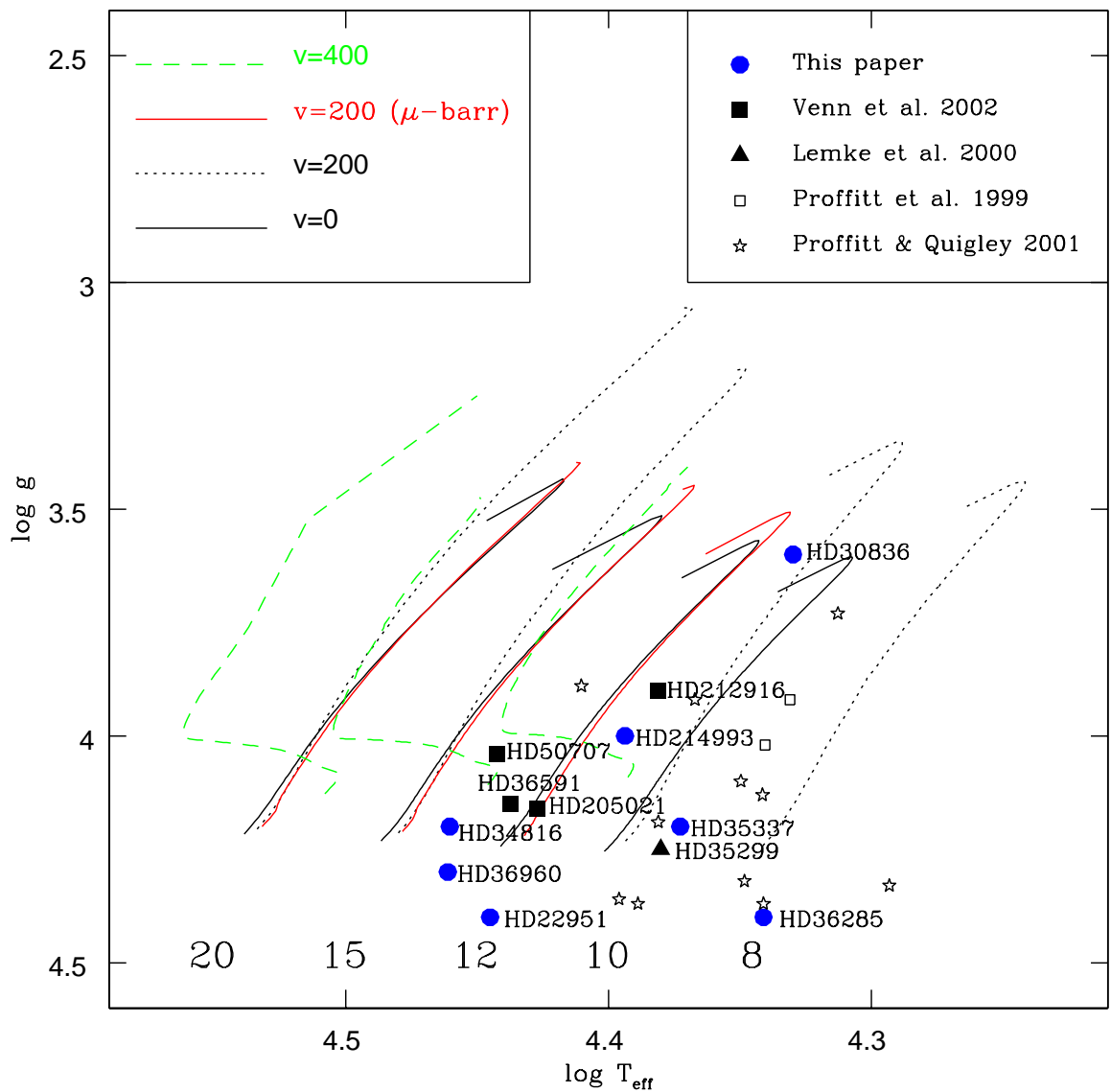}
\figcaption{Temperature vs gravity to determine spectroscopic masses.  
Models with low (black line) to moderate rotation rates (dotted black line)
and including $\mu$-barriers (red line) are from Heger \& Langer (2000).
The effects of very rapid rotation on main-sequence position and thus
spectroscopic mass can also be seen (green lines).   Stellar masses associated
with each model are labelled along the bottom of the figure.  
Only stars analysed in this paper and V02 are labelled.  
The temperatures of the stars have been scaled from their references
to the Cunha $\&$ Lambert (1994) scale (see V02 for details).
\label{gvsteff2}} 
\end{figure}

\begin{figure}[ht]
\epsscale{1}
\plotone{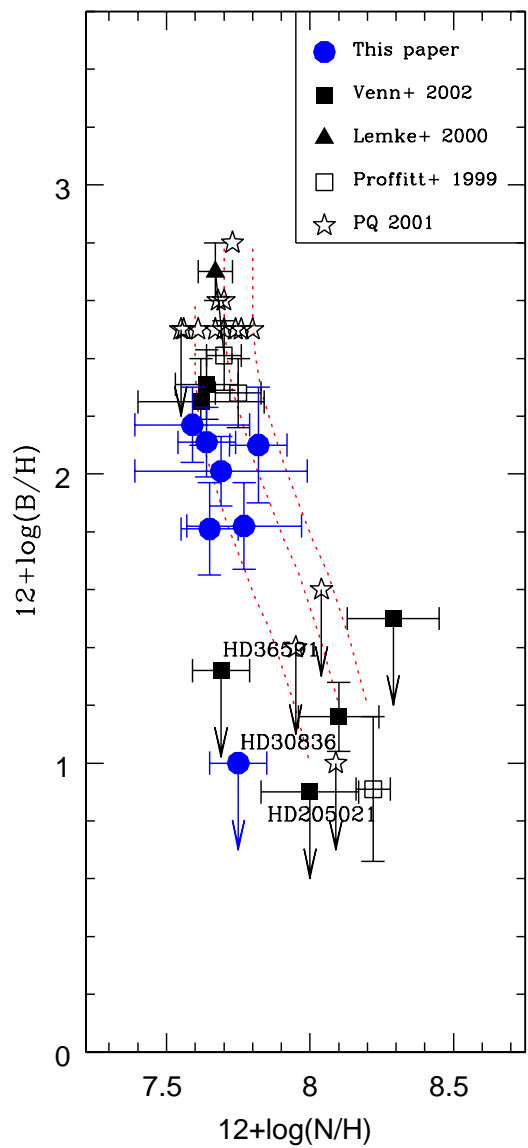}
\figcaption{Boron and nitrogen abundances in B-stars.  
Note that these abundances have been corrected from their
original references to a common temperature scale (see V02 for details). 
One rotating stellar evolution model is shown from Heger \& Langer (2000);
that for 12 M$_\odot$, V$_{\rm rot}$=200~\kms, and including the 
$\mu$-barrier (other masses and rotation rates have nearly identical 
predictions for the change in boron with nitrogen).   
The three dotted lines follow the predictions from
this model for three different initial nitrogen and boron 
abundances; $(12+$\logn$, 12+$\logb$) = (7.6, 2.6), 
(7.7,2.7), (7.8,2.8)$.
The x-axis and y-axis are scaled equally to emphasize the destruction
of boron before significant nitrogen enrichment. 
\label{modbn}
} 
\end{figure}

\begin{figure}[ht]
\epsscale{1}
\plotone{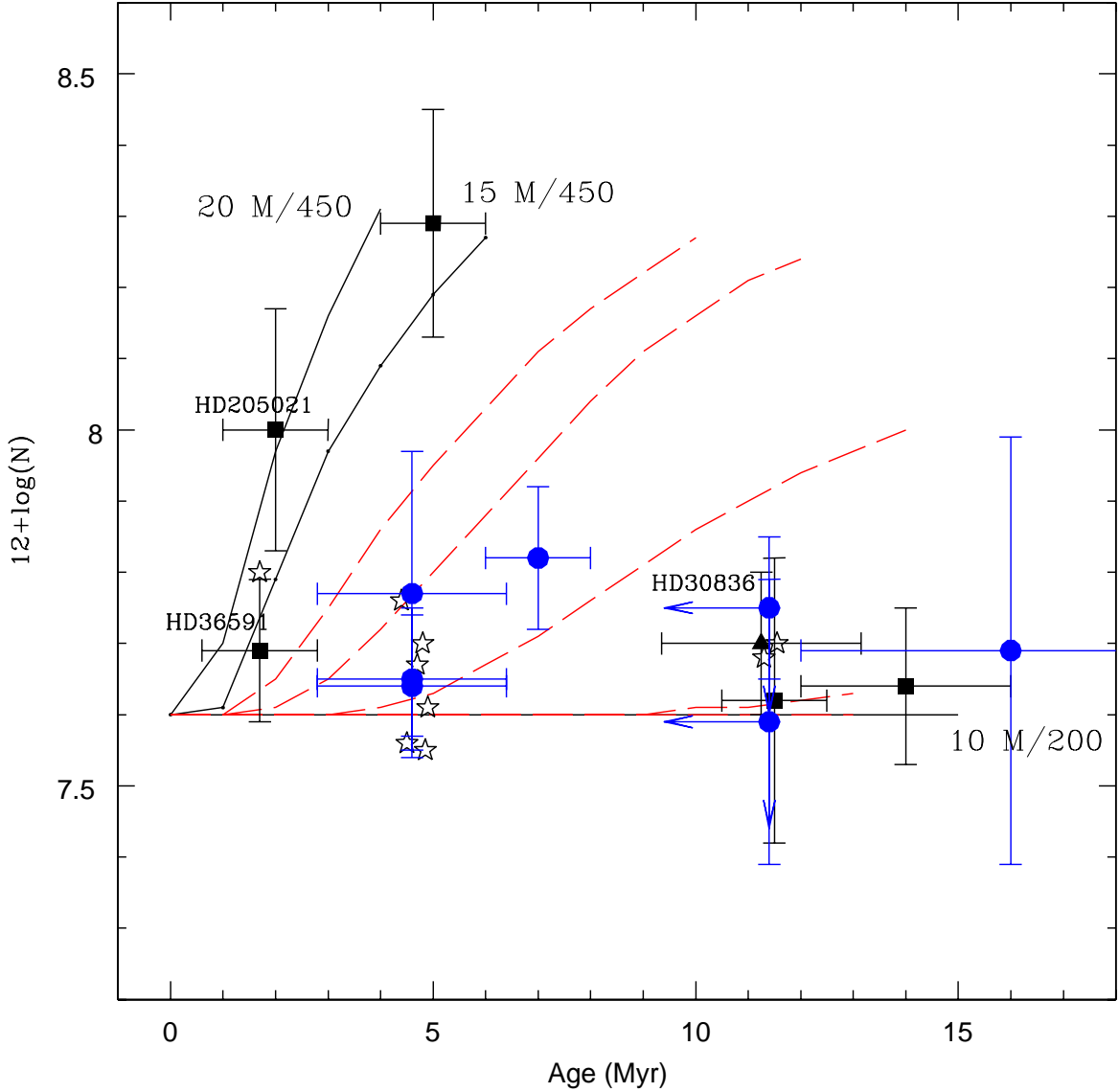}
\caption{Nitrogen enrichment as a function of age.
Data symbols the same as in Fig.~\ref{gvsteff2}.  
Rotating stellar evolution models are from Heger \& Langer (2000).
Red dashed lines represent 12$M_{\sun}$ models with 
various rotational velocities (0, 100, 200, 300, 450~\kms,
where the lines for the slowest rotation rates lie along the 
bottom of the plot).
Solid black lines represent 15 and 20$M_{\sun}$ models with the highest
rotation rate (450~\kms), and a 10$M_{\sun}$ model with a moderate
rotation rate (200~\kms). 
\label{nvsage}} 
\end{figure}

\begin{figure}[ht]
\epsscale{1}
\plotone{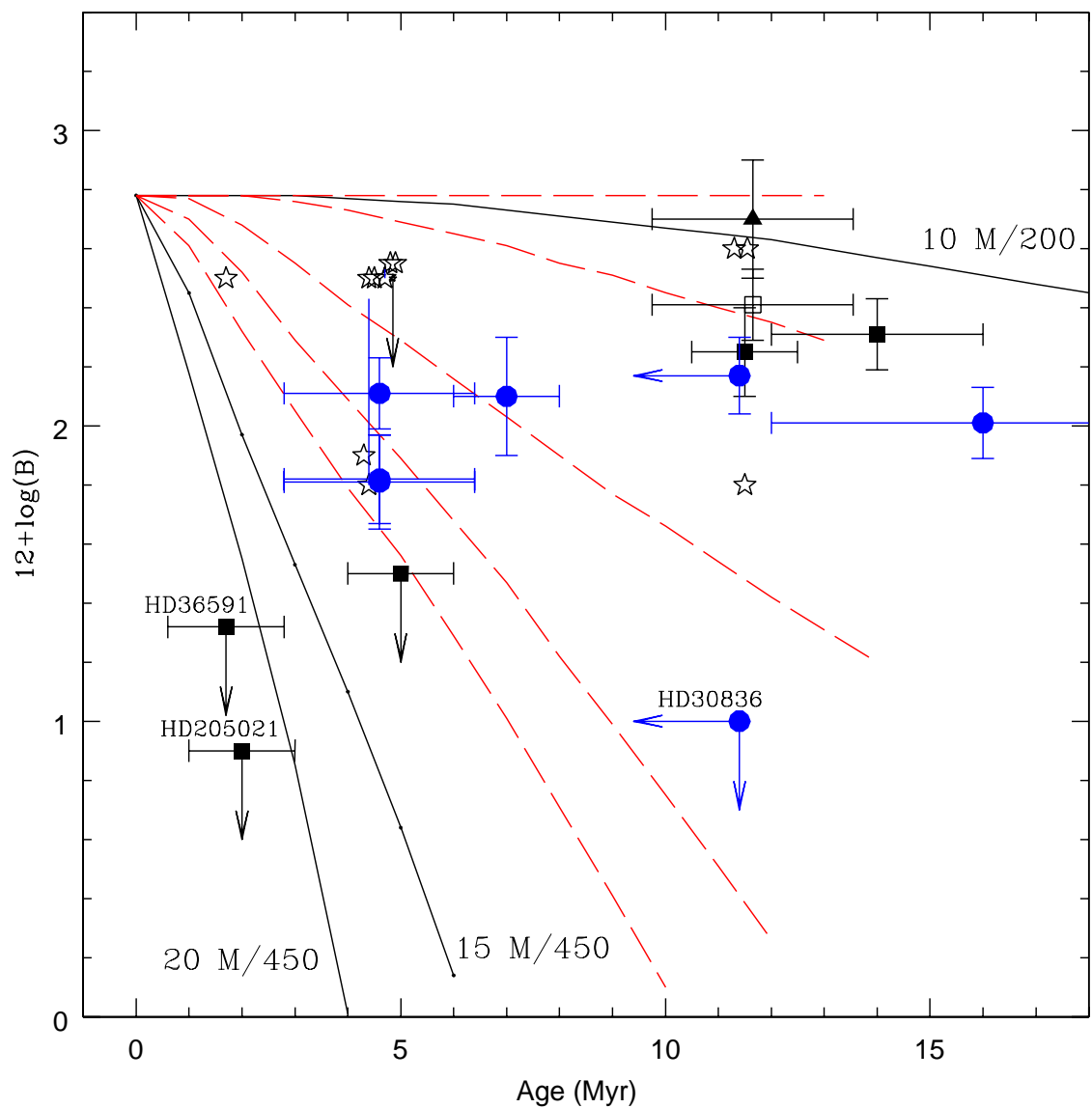}
\figcaption{Boron depletion as a function of age.   
Data symbols the same as in Fig.~\ref{gvsteff2}.  
Rotating stellar evolution models are from Heger \& Langer (2000).
Red dashed lines represent 12$M_{\sun}$ models with 
various rotational velocities (0, 100, 200, 300, 450~\kms).
Solid black lines represent 15 and 20$M_{\sun}$ models with the highest
rotation rate (450~\kms), and a 10$M_{\sun}$ model with a moderate
rotation rate (200~\kms). 
\label{bvsage}} 
\end{figure}

Those stars with larger boron depletions than predicted from models may 
indicate additional or more efficient mixing, 
but one of the major advantages of the data set in this paper is that 
the boron and nitrogen abundances have been determined for stars in 
clusters, thus stars which also have age constraints.   
In Figs. \ref{nvsage} and \ref{bvsage}, we show the nitrogen and boron 
abundances versus their stellar cluster ages, respectively.  
Ages are listed in Table~\ref{basics}.
The models shown in Figs. \ref{nvsage} and \ref{bvsage} 
(and also Fig.~\ref{gvsteff2}) 
include rotational mixing and $\mu$-barrier effects 
for several different stellar models across a span of rotational velocities 
(0-450~\kms) from 10 to 20~$M_{\sun}$ in order to illustrate the effects of 
these parameters on abundance.   

The predicted trend of boron-depletion and nitrogen-enrichment lends itself 
to certain constraints in the ability to detect boron if nitrogen is enhanced
and/or with age.  For example, boron is nearly undetectable after a nitrogen
enrichment of only 0.4~dex (based on Fig.~\ref{modbn}).   From
Fig.s~\ref{nvsage} and \ref{bvsage}, for a 12~$M_{\sun}$ star with 
v$_{rot}$=400~\kms, this time scale is no more than 10~Myr.  
Thus, boron's sensitivity to early 
mixing is far greater than that of nitrogen.

In comparing the observational stellar abundances to the model predictions,
the stellar data do show a large range in boron abundances,
from 0.9 $\ge$ 12+log(B/H) $\ge$ 2.9.  The stars with 12+log(B/H) $\ge$ 1.8
(which comprises 3/4 of the sample) {\it do not show significant nitrogen enrichments};
see Figs.~\ref{modbn} and \ref{nvsage}.
This suggests a range of a factor of 10 in boron {\it without} nitrogen enrichments, 
as predicted by stellar evolution models that include rotation.  The remaining 1/4 
of the sample, with 12+log(B/H) $\le$ 1.8, show a large range in nitrogen, some with 
no enrichments at all, i.e., HD~36591 and HD~30836.    These two stars are discussed,
independently, below.

In Fig.~\ref{bvsage} most of the stellar data is in good agreement with 
the model predictions (for $\sim$12 M$_\odot$ stars, and given the uncertainty
in the rotation rates).    Only two stars stand out, HD~36591 and HD~205021, 
because their ages are so young and boron abundances so low.
For HD\,36591, its young age and low boron abundance suggests 
a 20 $M_{\sun}$ model with the fastest rotation rate (450~\kms); 
see Fig.~\ref{bvsage}.   On the other hand, examination of Fig.~\ref{nvsage} 
suggests a 15 M$_\odot$ model with the fastest rotation rate.
However, its spectroscopic mass (from Fig.~\ref{gvsteff2}) 
is at most 13~$M_{\sun}$.  In order for the masses to agree, the 
temperature needs to be increased by $\sim$15-30\%.  
This result suggests that rotational mixing may be more 
efficient than currently modelled, at least at the highest 
rotation rates, and that the current models do not simultaneously
predict the boron abundance, nitrogen abundance, mass, and age for
this star.

The remaining two stars in our sample that stand out in either 
Fig.~\ref{modbn} or Fig.~\ref{bvsage} are HD\,30836 and HD\,205021. 
Both are severely depleted in boron, but have also been identified as 
spectroscopic binaries.   Binary mass
transfer is another way of depleting surface boron abundances
(Wellstein 2001; Langer, Wellstein, $\&$ Heger 2001).  
These stars have similar masses (10-12$M_{\sun}$ from Fig.~\ref{gvsteff2}) 
and orbital periods (10 days, Koch 1990).  
Models from Wellstein (2001) predict that mass transfer occurs in these 
systems shortly after core hydrogen burning in the primary star. 
While binary mass transfer can produce boron depletions, only rotational 
mixing predicts boron depletion {\it without} nitrogen enrichments.
HD\,205021 shows a nitrogen enrichment (\logn~=8.0) in addition to 
its depleted boron abundance.  We notice in 
Fig.~\ref{nvsage} that the nitrogen enrichment and age for this star
implies a mass of $\sim$20 M$_\odot$ and the highest rotation
rate (450 \kms).   The same result is found examining its low boron 
abundance and age in Fig.~\ref{bvsage}.   However, its spectroscopic
mass is only $\sim$12 M$_\odot$ (Fig.~\ref{gvsteff2}).   
Thus this star does not match the rotating model predictions, but 
could be affected by binary mass transfer.

HD\,30836 shows no nitrogen enrichment (\logn=7.79; Table~3), thus rotational 
mixing {\it uniquely} describes its abundance pattern.
Ironically, we only have an upper-limit to the boron abundance in HD\,30836,
however that upper-limit is extremely low.  We also only have an upper-limit
to its age from membership in the Orion OB1 association (the sub-association
is unclear for this star).   From the rotating models, the upper-limit age 
and upper-limit boron abundance suggest that HD\,30836 has a mass of 
12~$M_{\sun}$ with a rotational velocity between 200 and 300~\kms (see
Fig.~\ref{bvsage}).  However, the spectroscopic mass (see Fig.~\ref{gvsteff2})
for this star is closer to 10~$M_{\sun}$.  Adopting 10~$M_{\sun}$ 
in Fig.~\ref{bvsage} would require a rotational velocity over 450~\kms, 
higher than the theoretical upper limit.  Similarly, adopting a 12~$M_{\sun}$ 
in Fig.~\ref{gvsteff2} would require an increase of nearly 3000~K 
(approximatly 15\%) in temperature.
Thus, the nitrogen abundance and upper-limits to age and boron abundance 
make HD\,30836 another interesting star for comparing with the rotating
stellar models, regardless of its binarity.   This star, along with  
HD\,36951 (and possibly HD\,205021) are best fit by models with masses 
significantly larger than their spectroscopic masses.   This 
suggests that rotational mixing is more efficient than currently 
modelled at the highest rotation rates.

\section{Conclusions}

Detailed abundances of boron have been determined in a careful selection
of seven sharp-lined B-stars in young clusters to test massive star evolution 
scenarios that include rotational mixing effects.   The seven stars in this 
analysis show moderate-to-severe depletions of boron (1/4 to 1/50 solar!) 
{\it without} significant nitrogen enrichments.   This is a unique prediction
of the massive star evolution models that include rotational mixing 
(Heger $\&$ Langer 2000), because shallow mixing will destroy surface boron
abundances before deep mixing bring CN-cycled gas to the surface.    
Only three stars deviate from the model predictions, 
HD\,30836, HD\,36591, and HD\,205021.   In all three cases, their
spectroscopic masses are smaller than predicted from the rotating
evolution models (i.e., the masses required to explain their low
boron abundances and young ages).   
These three stars appear to indicate that rotational mixing is 
more efficient than currently modelled at the highest rotation rates.
These results are consistent with previously published results 
(Venn \etal\ 2002), but approximately doubles the sample size of stars 
with both moderate and severe boron depletions.

\acknowledgements
We would like to thank Katia Cunha for help in analysing HD\,35299,
and many productive and enjoyable conversations.
This research was supported by NASA grant HST GO-09437. 
Space Telescope Science Institute is operated by the Association of Universities 
for Research in Astronomy, Inc. under NASA contract NAS 5-26555.
DLL thanks the Robert A. Welch Foundation of Houston, Texas for their support.

\begin{deluxetable}{lccccccl}
\tabletypesize{\footnotesize}
\tablecaption{Basic Parameters \label{basics}}
\tablewidth{0pt}
\tablehead{
\colhead{} &\colhead{} &\colhead{} &\colhead{} &\colhead{} &\colhead{} &\colhead{Assoc. Age} &\colhead{}\\
\colhead{Star} &\colhead{Name} &\colhead{Sp. Type} &\colhead{Variable} &\colhead{Binary} &\colhead{OB Assoc.} &\colhead{(Myr)} &\colhead{Ref.} 
}
\startdata
HD 22951        &o Per         &B0.5V  &\nodata     &Visual? (check)	&Per 0B2  &12-20                &1,3,4 \\ 
HD 30836 	&$\pi^4$ Ori   &B2III  &\nodata     &Spect.	        &Ori OB1  &$\leq$11.4           &2,5,6 \\
HD 34816 	&$\lambda$ Lep &B0.5IV &\nodata     &\nodata	        &Ori OB1  &$\leq$11.4           &2 \\
HD 35337 	&8 Lep         &B2IV   &\nodata     &\nodata            &Ori OB1c &4.6$^{+1.8}_{-2.1}$  &2,6 \\
HD 36285 	&\nodata       &B2IV-V &\nodata     &\nodata            &Ori OB1c &4.6$^{+1.8}_{-2.1}$  &2,6 \\
HD 36960 	&\nodata       &B0.5V  &Variable?   &Visual             &Ori OB1c &4.6$^{+1.8}_{-2.1}$  &2,7,8 \\
HD 214993 	&12 Lac        &B2III  &$\beta$ Cep &\nodata            &Lac OB1  &6-8                  &1,9 \\
\enddata
\tablerefs{ (1)de Zeeuw $\&$ Brand 1985 (2)Brown, de Geus, $\&$ de Zeeuw 1994 (3)Klockova $\&$ Kopylov 1985 (4)Petrova $\&$ Orlov 1999 (5)Batten, Fletcher, $\&$ Mann 1978 (6)Abt $\&$ Cardona 1984 (7)Morrell $\&$ Levato 1991 (8)Lindroos 1986 (9)Struve 1951}
\end{deluxetable}

\clearpage
\begin{deluxetable}{lccrrr}
\tabletypesize{\footnotesize}
\tablecaption{HST STIS Observing Information for Galactic B-stars\label{obs}}
\tablewidth{0pt}
\tablehead{
\colhead{Star} & \colhead{V} & \colhead{Grat/Slit} & 
\colhead{Exposure(s)} & \colhead{Date} & \colhead{S/N}
}
\startdata
HD 22951  	&  4.98 & E230M         &   1707 at $\lambda_c$2124 &2003 Mar 25 & 50 \\
     		&       & 0.2x0.05ND    &  2634 at $\lambda_c$2124 & \\ 
	     	&       &	        &  2773 at $\lambda_c$1978 & \\ 
	     	&       &       	&  2760 at $\lambda_c$2269 & \\
HD 30836	& 3.67  & E230M		&  856 at $\lambda_c$2124 &2003 Apr 6 & 65 \\
		&	& 0.2x0.05ND 	&  720 at  $\lambda_c$2124 & \\
		&	&		&  1200 at  $\lambda_c$1978 & \\
		&	&		&  1091 at  $\lambda_c$2269 &  \\
		&	&		&  720 at  $\lambda_c$2124 &2003 Aug 2 & \\
		&	&		&  600 at  $\lambda_c$2124 & \\
		&	&		&  1200 at  $\lambda_c$1978 & \\
		&	&		&  1200 at  $\lambda_c$2269 &  \\
HD 34816	& 4.27  & E230M		&  855 at  $\lambda_c$2124 &2003 Apr 25 & 80 \\ 
		&	& 0.2x0.05ND	&  720 at  $\lambda_c$2124 & \\
		&	&		&  1200 at  $\lambda_c$1978 &  \\
		&	&		&  1091 at $\lambda_c$2269 & \\
HD 35337	& 5.22	& E230M		&  3036 at  $\lambda_c$2124 &2003 Apr 24 & 100 \\
		&	& 0.2x0.05ND	&  2166 at  $\lambda_c$1978 & \\
		&	&		&  1347 at  $\lambda_c$2269 &  \\
HD 36285	& 6.32	& E230H		&  572 at  $\lambda_c$2013 &2003 Apr 28 & 280 \\
		&	& 31x.0.05NDA	&  1080 at  $\lambda_c$2013 &  \\
		&	&		&  1260 at  $\lambda_c$2013 &  \\
		&	&		&  1272 at  $\lambda_c$2013 &  \\
		&	&		&  1200 at  $\lambda_c$2013 &  \\
		&	&		&  624 at  $\lambda_c$2013 &  \\
		&	&		&  1410 at  $\lambda_c$2013 &  \\
		&	&		&  1410 at  $\lambda_c$2013 &  \\
		&	&		&  1410 at  $\lambda_c$2013 &  \\
		&	&		&  1388 at  $\lambda_c$2013 &  \\
HD 36960	& 4.78	& E230M		&  1661 at  $\lambda_c$2124 &2003 Apr 27 & 90 \\
		&	& 0.2x0.05ND	&  1560 at  $\lambda_c$1978 &  \\
		&	&		&  720 at  $\lambda_c$2269 &  \\
HD 214993	& 5.23	& E230M		&  2160 at  $\lambda_c$2124 &2003 Jul 7 & 40 \\
		&	& 0.2x0.05ND	&  2640 at  $\lambda_c$1978 &2003 Jul 8 &  \\
		&	&		&  1800 at  $\lambda_c$2269 &  \\
\enddata
\end{deluxetable}
\clearpage

\begin{deluxetable}{lccccc|cc}
\tabletypesize{\footnotesize}
\tablecaption{Stellar Parameters from Literature \label{param2}}
\tablewidth{0pt}
\tablehead{
\colhead{} &\colhead{GL92-3.4$\%$} &\colhead{CL94} &\colhead{} &
\colhead{NLTE} &\colhead{NLTE} &\colhead{Corrected} &\colhead{Corrected} \\
\colhead{Star} &\colhead{\teff} &\colhead{\teff} &\colhead{\logg} &
\colhead{log(N/H)} &\colhead{log(O/H)} &\colhead{log(N/H)} &\colhead{log(O/H)}   \\
\colhead{} &\colhead{(K)} &\colhead{(K)} &\colhead{} &
\colhead{} &\colhead{} &\colhead{} &\colhead{} 
}
\startdata
HD 22951  &27870   &\nodata &4.40 &7.69  &8.45    &7.69  &8.42      \\ 
HD 30836  &21370   &\nodata &3.60 &7.79  &\nodata &7.75  &\nodata   \\
HD 34816  &28870   &\nodata &4.20 &7.66  &8.75    &7.59  &8.67      \\
HD 35337  &23590   &\nodata &4.20 &7.65  &8.56    &7.64  &8.55      \\
HD 36285  &\nodata &21930   &4.40 &7.77  &8.79    &7.77  &8.55      \\
HD 36960  &28940   &28920   &4.30 &7.72  &8.88    &7.65  &8.80      \\
HD 214993 &24760   &\nodata &4.00 &7.80  &8.78    &7.82  &8.83     \\
\enddata
\tablecomments{Corrections made to the nitrogen and oxygen abundances 
are based on both the reduced GL92 temperature scale and an adjustment from the 
less heavily line-blanketed Gold models to those of Kurucz, as discussed by CL94.
Adjustments are made according 
to the errors found in Table 9 of GL92 ($\Delta$ values) and Table 10 of CL94.}
\end{deluxetable}

\begin{deluxetable}{lcccccc}
\tabletypesize{\footnotesize}
\tablewidth{0pc}
\tablecaption{Iron-group Abundance Results\label{mabund}}
\tablehead{
\colhead{$\lambda$} &\colhead{} &\colhead{[M/H]} &\colhead{[M/H]} &\colhead{[M/H]} &\colhead{[M/H]} &\colhead{M/H]} \\ 
\colhead{(\AA)} &\colhead {Element(s)} &\colhead{HD\,22951} & \colhead{HD\,34816} & \colhead{HD\,35337} & \colhead{HD\,36285} & \colhead{HD\,36960}
}
\startdata
2048.92..... & \ion{Mn}{3}				&{\it+0.26}	&-0.09 		&+0.13		 &-0.04		 &-0.06 \\
2049.37..... & \ion{Fe}{3}+\ion{Mn}{3}   		&-0.20		&{\it+0.00}	&-0.08		 &-0.05		 &-0.10 \\
2049.66..... & \ion{Mn}{3}               		&-0.51		&-0.43	 	&{\it+0.38}      &{\it+0.16}	 &-0.31 \\
2050.74..... & \ion{Fe}{3}               	 	&{\it-0.90} 	&-0.25 		&+0.01 		 &-0.35	 	 &-0.12 \\
2051.85..... & \ion{Fe}{3}+\ion{Fe}{4}  		&{\it-0.05}	&-0.15 		&-0.09 		 &-0.10		 &-0.15 \\
\textbf{2052.27...} & \ion{Fe}{3}              		&-0.60		&-0.41 		&-0.17 		 &-0.35		 &-0.27 \\
2053.52..... & \ion{Fe}{3}              	  	&{\it-1.00} 	&{\it-0.71}	&{\it-0.43}	 &-0.38		 &{\it-0.45} \\
\textbf{2054.56...} & \ion{Fe}{3}x3			&-0.37		&-0.23 		&-0.13	 	 &-0.04		 &-0.10  \\
2055.86..... & \ion{Fe}{3}				&-0.17		&{\it+0.02}	&+0.03 		 &-0.21		 &{\it+0.14} \\
2056.16..... & \ion{Fe}{3}				&-0.59		&{\it+0.13}	&-0.20 		 &-0.25		 &{\it+0.15} \\
2057.07..... & \ion{Fe}{3}				&-0.54		&{\it 0.00}	&-0.22 		 &{\it-0.46}	 &{\it+0.25} \\
\textbf{2057.93...} & \ion{Fe}{3}			&-0.49		&-0.10 		&-0.27 		 &-0.40		 &-0.09 \\
2058.21..... & \ion{Fe}{3}				&{\it-0.11}	&{\it+0.15}	&-0.05 		 &-0.08		 &{\it+0.25} \\
2058.57..... & \ion{Fe}{3}				&-0.35		&{\it+0.35}	&{\it+0.17} 	 &-0.10		 &{\it+0.35} \\
2059.67..... & \ion{Fe}{3}				&{\it-0.82}	&{\it+0.07}	&{\it+0.15}	 &-0.20		 &-0.01 \\
2063.40..... & \ion{Mn}{3}				&-0.42		&-0.54 		&{\it+0.16}	 &-0.14		 &-0.34 \\
\textbf{2066.40...} & \ion{Mn}{3}x2+\ion{Ni}{3}		&-0.54		&-0.54 		&-0.14 		 &-0.28		 &-0.39 \\
2068.26..... & \ion{Fe}{3}				&-0.57		&{\it+0.38}	&{\it+0.25} 	 &{\it+0.05}	 &{\it+0.38} \\
2068.99..... & \ion{Fe}{3}+\ion{Mn}{3}+\ion{Cr}{3}	&-0.70 		&{\it-0.70}	&-0.10	 	 &-0.05		 &{\it-0.50} \\
\textbf{2069.82...} & \ion{Fe}{3}+\ion{Mn}{3}		&-0.20		&-0.20 		&-0.09 		 &-0.09		 &-0.15 \\
2070.56..... & \ion{Fe}{3}				&-0.40		&{\it+0.08}	&-0.14 		 &-0.36		 &{\it+0.15} \\
2070.98..... & \ion{Fe}{3}x3				&-0.55		&-0.48 		&{\it-0.32} 	 &-0.32		 &-0.33 \\
\textbf{2073.35...} & \ion{Fe}{3}+\ion{Mn}{3}		&-0.30		&-0.28 		&-0.05 		 &-0.13		 &-0.30 \\
2074.23..... & \ion{Fe}{3}				&-0.62		&{\it-0.65}	&{\it-0.32} 	 &{\it-0.55}	 &-0.40 \\
2076.32..... & \ion{Fe}{3}				&-0.63		&{\it-0.74}	&-0.05 		 &{\it-0.50}	 &{\it-0.72} \\
\textbf{2077.36...} & \ion{Mn}{3}+\ion{Co}{3}		&-0.23		&-0.53 		&+0.04 		 &-0.13		 &-0.33 \\
2077.74..... & \ion{Fe}{3}+\ion{Fe}{4}			&{\it+0.12} 	&{\it+0.10}	&{\it+0.15}	 &{\it+0.01}	 &{\it+0.10} \\
2078.08..... & \ion{Fe}{3}+\ion{Mn}{3}			&-0.30		&-0.40 		&{\it-0.43} 	 &{\it-0.50}	 &-0.43 \\
2079.00..... & \ion{Fe}{3}x4				&-0.60		&{\it+0.25}	&+0.09 		 &-0.45		 &{\it+0.20} \\
\textbf{2080.22...} & \ion{Fe}{3}			&-0.30		&-0.38 		&-0.10 		 &-0.25		 &-0.29 \\
\textbf{2081.08...} & \ion{Mn}{3}x2+\ion{Co}{3}	        &-0.25		&-0.13 		&-0.06 		 &-0.28		 &-0.23 \\
\textbf{2082.38...} & \ion{Fe}{3}			&-0.25		&-0.35 		&+0.00 		 &-0.10		 &-0.36 \\
\textbf{2083.55...} & \ion{Fe}{3}			&-0.20		&-0.20 		&+0.05 		 &-0.25		 &-0.07 \\
2084.36..... & \ion{Fe}{3}x3+\ion{Mn}{3}x2		&-0.60		&-0.43 		&-0.15 		 &{\it-0.65}	 &-0.37 \\
2084.93..... & \ion{Fe}{3}x2				&-0.40		&-0.30		&{\it+0.15}	 &-0.20		 &-0.15 \\
\textbf{2085.84...} & \ion{Fe}{3}+\ion{Cr}{3}		&-0.25		&-0.15 		&+0.02 		 &-0.14		 &-0.15 \\
2087.15..... & \ion{Fe}{3}				&{\it-1.05} 	&-0.31 		&+0.10	 	 &-0.37		 &-0.38 \\
2087.93..... & \ion{Fe}{3}				&{\it-0.80}	&-0.51 		&+0.13 		 &-0.40		 &-0.44 \\
\textbf{2089.12...} & \ion{Fe}{3}			&-0.65		&-0.03 		&+0.01 		 &-0.45		 &+0.02 \\
2090.16..... & \ion{Fe}{3}x4+\ion{Mn}{3}x3		&{\it-1.10} 	&-0.45 		&+0.05 		 &-0.15		 &{\it-0.45} \\
2091.35..... & \ion{Fe}{3}x2				&-0.65		&-0.46 		&{\it-0.32} 	 &{\it-0.48}	 &-0.40 \\

2092.97..... & \ion{Fe}{3}				&-0.20		&{\it+0.00}	&-0.05 		 &-0.22		 &{\it+0.15} \\
2093.51..... & \ion{Fe}{3}				&-0.25		&{\it-0.58}	&-0.08 		 &-0.25		 &-0.13 \\
2095.66..... & \ion{Fe}{3}x3				&-0.25		&-0.44 		&-0.15 		 &-0.05		 &-0.20 \\
2096.42..... & \ion{Fe}{3}				&-0.60		&-0.40 		&-0.14	 	 &-0.23		 &-0.05 \\  
2099.30..... & \ion{Fe}{3}x2				&-0.50		&{\it-0.55}	&-0.15 		 &-0.45		 &-0.42 \\  
\textbf{2101.04...} & \ion{Fe}{3}+\ion{Mn}{3}		&-0.50		&-0.30 		&-0.07 		 &-0.18		 &-0.20 \\   
2103.74..... & \ion{Fe}{3}x4				&{\it-1.00} 	&-0.48 		&{\it-0.40} 	 &{\it-0.60}	 &-0.39 \\  
2104.96..... & \ion{Fe}{3}x2+\ion{Cr}{3}+\ion{Ni}{3}	&-0.32		&-0.20 		&{\it+0.15}	 &{\it+0.03}	 &-0.18 \\   
2105.59..... & \ion{Cr}{3}				&{\it-0.15}	&-0.19 		&-0.12 		 &-0.10		 &-0.22 \\   
2107.34..... & \ion{Fe}{3}				&-0.45		&-0.20 		&{\it-0.36} 	 &-0.40		 &+0.00 \\   
\textbf{2108.64...} & \ion{Fe}{3}+\ion{Mn}{3}		&-0.42		&-0.43 		&-0.05 		 &-0.07		 &-0.30 \\   
2111.80..... & \ion{Fe}{3}x2				&-0.30 		&-0.50 		&{\it-0.43} 	 &-0.45		 &-0.43 \\   
2113.34..... & \ion{Fe}{3}x2+\ion{Mn}{3}		&{\it+0.00}	&-0.50 		&-0.04 		 &-0.30		 &-0.18 \\   
\textbf{2113.83...} & \ion{Fe}{3}x2+\ion{Cr}{3}		&-0.40		&-0.35 		&+0.02 		 &-0.23		 &-0.10 \\  
2114.34..... & \ion{Fe}{3}x2+\ion{Cr}{3}		&{\it-0.15}	&-0.25 		&+0.00 		 &-0.11		 &-0.10 \\   
2114.88..... & \ion{Cr}{3}x2				&{\it-0.10} 	&{\it+0.00}	&{\it+0.06} 	 &-0.05		 &{\it+0.08} \\  
2116.59..... & \ion{Fe}{3}x2				&{\it-0.95} 	&{\it-0.86}	&+0.07 		 &{\it-0.48}	 &{\it-0.72} \\  
\textbf{2117.55...} & \ion{Cr}{3}			&-0.20 		&-0.27		&-0.14 		 &-0.20		 &-0.02 \\  
\textbf{2118.49...} & \ion{Fe}{3}x2+\ion{Cr}{3}		&-0.45		&-0.40 		&-0.05	 	 &-0.25		 &-0.30 \\  
\textbf{2120.77...} & \ion{Fe}{3}x4+\ion{Fe}{4}		&-0.53		&-0.20 		&-0.07 		 &-0.07		 &-0.15 \\  
\textbf{2123.59...} & \ion{Fe}{3}x3+\ion{Cr}{3}		&-0.50		&-0.40 		&-0.02 		 &-0.05		 &-0.27 \\  
2124.16..... & \ion{Fe}{3}x5+\ion{Co}{3}		&{\it-0.80}	&{\it-0.70}	&{\it-0.40} 	 &-0.31		 &{\it-0.65} \\  
2125.18..... & \ion{Fe}{3}x5+\ion{Fe}{4}+\ion{Mn}{3}	&-0.55		&-0.45 		&{\it-0.39} 	 &-0.33		 &-0.35 \\  
\textbf{2126.14...} & \ion{Mn}{3}			&-0.28		&-0.43 		&+0.00 		 &-0.04		 &-0.39 \\  
\textbf{2129.68...} & \ion{Fe}{3}x3			&-0.20		&-0.27 		&-0.08 		 &-0.04		 &-0.23 \\   
2134.83..... & \ion{Fe}{3}x2				&{\it-0.08} 	&-0.04 		&{\it+0.29} 	 &{\it+0.03}	 &{\it+0.12} \\   
2136.36..... & \ion{Fe}{3}x2				&-0.56 		&{\it-0.74}	&{\it-0.49} 	 &{\it-0.65}	 &{\it-0.65} \\   
Average....		&				&-0.42		&-0.33 		&-0.07 		 &-0.22		 &-0.23 \\   
1 $\sigma$............	&				&+0.15		&+0.16 		&+0.15		 &+0.14		 &+0.15 \\   
\enddata
\tablecomments{
Results that fall $\geq 2\sigma$ from the mean  are {\it italicized}
and not included in the final average or 1 $\sigma$ values.  
Wavelengths in \textbf{bold} are those that show consistent results 
across all five stars (none fall outside $2\sigma$).  }
\end{deluxetable}

\begin{deluxetable}{lccccc}
\tabletypesize{\footnotesize}
\tablecaption{Atmospheric Analysis Results \label{results}}
\tablewidth{0pt}
\tablehead{
\colhead{} &\colhead{$v$\,sin\,$i$} &\colhead{$\xi$} &\colhead{} &\colhead{LTE} &\colhead{NLTE\tablenotemark{1}}  \\
\colhead{Star} &\colhead{\kms} &\colhead{\kms} &\colhead{\logf} &\colhead{\logb} &\colhead{\logb}  
}
\startdata
HD 22951 	&23 &1 &7.03$\pm$0.10 		&2.01$\pm$0.12 	& 2.01   \\
HD 30836 	&43 &4 &{\it 7.25}$\pm$0.25 	&$<$1.2         & $<$1.0 \\
HD 34816 	&35 &3 &7.12$\pm$0.11 		&2.17$\pm$0.13 	& 2.17   \\
HD 35337 	&15 &1 &7.38$\pm$0.10 		&2.21$\pm$0.12 	& 2.11   \\
HD 36285 	&10 &2 &7.23$\pm$0.09 		&2.02$\pm$0.15 	& 1.82   \\
HD 36960 	&33 &3 &7.22$\pm$0.10 		&1.81$\pm$0.16 	& 1.81   \\
HD 214993 	&39 &5 &{\it 7.25}$\pm$0.25 	&2.2$\pm$0.2    & 2.1 \\

\enddata
\tablecomments{Macroturbulence values were set by the instrumental broadening;
9 and 4~\kms for observations with the E230M and E230H gratings, respectively.}
\tablenotetext{1}{NLTE corrections were mading according to the grid used in V02 (their Fig.~9)}
\end{deluxetable}

\begin{deluxetable}{lccccc}
\tabletypesize{\footnotesize}
\tablewidth{0pc}
\tablecaption{Boron Abundance Uncertainties\label{errors}}
\tablehead{
\colhead{} &\colhead{HD22951} &\colhead{HD34816} &\colhead{HD35337} &\colhead{HD36285} &\colhead{HD36960} \\
\colhead{Parameters} &\colhead{$\Delta$\logb)} &\colhead{$\Delta$\logb} &\colhead{$\Delta$\logb} &\colhead{$\Delta$\logb} &\colhead{$\Delta$\logb}
}
\startdata
$\Delta$T$_{eff}$ = $\pm$750 K	&$\pm$0.07 &$\pm$0.07 	&$\pm$0.06 &$\mp$0.05 &$\pm$0.12 \\
$\Delta$ log $g$ = $\pm$0.2	&$\pm$0.08 &$\pm$0.07 	&$\pm$0.10 &$\pm$0.13 &$\pm$0.04 \\
$\Delta \xi$ = $\pm$ 1 \kms	&$\mp$0.01 &$\mp$0.03 	&$\mp$0.02 &$\mp$0.02 &$\mp$0.01 \\
$\Delta \xi_{ma}$ = $\pm$ 2 \kms&$\pm$0.00 &$\pm$0.02 	&$\pm$0.00 &$\pm$0.00 &$\pm$0.00 \\
Shift continuum $\pm$ 1$\%$	&$\mp$0.05 &$\mp$0.08 	&$\mp$0.03 &$\mp$0.04 &$\mp$0.09 \\
$\Delta$V$_{rad}$ = $\pm$ 2 \kms&$\pm$0.00 &$\pm$0.01 	&$\pm$0.02 &$\pm$0.02 &$\pm$0.01 \\
Total Error\tablenotemark{1}	&$\pm$0.12 &$\pm$0.13	&$\pm$0.12 &$\pm$0.15 &$\pm$0.16 \\

\enddata
\tablenotetext{1}{Total error is taken as the sum of squares of all of the uncertainties.}
\end{deluxetable}

\begin{deluxetable}{lccc}
\tabletypesize{\footnotesize}
\tablecaption{Atmospheric Parameters from Lyubimkov \etal\ (2002) \label{L02param}}
\tablewidth{0pt}
\tablehead{
\colhead{} & \colhead{Lyubimkov02} &
\colhead{$\Delta$} &\colhead{$\Delta$} \\[.2ex] 
\colhead{Star} &\colhead{\teff \ \ \ \logg}  &  
\colhead{[Fe/H]} &\colhead{[B/H]} 
}
\startdata
HD 22951 	&26300  4.00 & 0.26 & 0.18 \\
HD 34816 	&27900  4.20 & 0.10 & 0.11 \\
HD 35337 	&22300  4.00 & 0.12 & 0.06 \\
HD 36285 	&21300  4.25 & 0.02 & 0.05 \\
HD 36960 	&27500  4.15 & 0.15 & 0.13 \\
\enddata
\tablecomments{$\Delta$ = log(X/H)$_{adopted}$ - log(X/H)$_{L02}$}
\end{deluxetable}

\clearpage

\clearpage

\clearpage

\clearpage

\clearpage

\clearpage

\clearpage

\clearpage

\clearpage

\clearpage

\end{document}